\documentclass[aps, prx, twocolumn,show pacs,show keys,superscript address,superscript reference,floatfix]{revtex4-2}
\usepackage[colorlinks=true,citecolor=blue,linkcolor=blue,urlcolor=blue]{hyperref}
\usepackage[sort&compress]{natbib}
\usepackage{graphicx,times}
\usepackage{color}
\usepackage{amssymb}
\newcommand{\red}[1]{\textcolor[rgb]{1.00,0.00,0.00}{#1}}
\usepackage{subeqnarray}
\usepackage{float}
\usepackage{bbm}
\usepackage{bm}
\usepackage{diagbox}
\usepackage{makecell}
\usepackage{amsmath}
\usepackage{balance}

\DeclareMathAlphabet\mathbfcal{OMS}{cmsy}{b}{n}
\newcommand{\ket}[1]{\ensuremath{\left|{#1}\right\rangle}}

\newcommand{\ketbra}[2]{\vert #1 \rangle \langle #2 \vert}

\begin{document}
\title{Microwave Quantum Memcapacitor Effect}
\author{Xinyu~Qiu}
\thanks{These authors contributed equally}
\affiliation{Physics Department, Shanghai University, 200444 Shanghai, China}

\author{Shubham~Kumar}
\thanks{These authors contributed equally}
\affiliation{Kipu Quantum,  Greifswalderstrasse 226, 10405 Berlin, Germany}

\author{Francisco~A.~C\'ardenas-L\'opez}
\affiliation{Forschungszentrum J\"ulich GmbH, Peter Gr\"unberg Institute, Quantum Control (PGI-8), 52425 J\"ulich, Germany} 

\author{Gabriel~Alvarado~Barrios}
\email{phys.gabriel@gmail.com}
\affiliation{Kipu Quantum,  Greifswalderstrasse 226, 10405 Berlin, Germany}

\author{Enrique~Solano}
\affiliation{Kipu Quantum,  Greifswalderstrasse 226, 10405 Berlin, Germany}

\author{Francisco~Albarr\'an-Arriagada}
\email{ francisco.albarran@usach.cl}
\affiliation{Departamento de F\'isica, CEDENNA, Universidad de Santiago de Chile (USACH), Avenida V\'ictor Jara 3493, 9170124, Santiago, Chile}

\begin{abstract}

\begin{center}
\textbf{Abstract}
\end{center}

Developing the field of neuromorphic quantum computing necessitates designing scalable quantum memory devices. Here, we propose a superconducting quantum memory device in the microwave regime, termed as a microwave quantum memcapacitor. It comprises two linked resonators, the primary one is coupled to a Superconducting Quantum Interference Device, which allows for the modulation of the resonator properties through external magnetic flux. The auxiliary resonator, operated through weak measurements, provides feedback to the primary resonator, ensuring stable memory behaviour. This device operates with a classical input in one cavity while reading the response in the other, serving as a fundamental building block toward arrays of microwave quantum memcapacitors. We observe that a bipartite setup can retain its memory behaviour and gains entanglement and quantum correlations. Our findings pave the way for the experimental implementation of memcapacitive superconducting quantum devices and memory device arrays for neuromorphic quantum computing.

\end{abstract}

\maketitle

\section{Introduction}
\label{sec.1}

Neuromorphic computing has emerged as a promising avenue for energy-efficient and advanced computing systems~\citep{Markovic2020NatRevPhys}, utilizing nonlinear devices with memory properties such as phase-change memory, transistors, spintronic devices, and memory devices~ \cite{Asapu.2019} to achieve heightened computational capabilities. Memristors, as nonlinear resistors, can be well described by Kubo's response theory~ \cite{Millar.1951,Kubo1957JPhysSocJpn}, where one characteristic feature is the pinched hysteresis loop in their input-output relation, which can be associated to memory properties~\cite{Biolek2014}. In 1971, L. Chua introduced the memristor concept as a theoretical fourth fundamental circuit element~\cite{IEEE.1971}. Its experimental realization was later confirmed by HP Labs in 2008~\cite{Nat.2008}. However, the precise existence of the ideal memristor, as postulated by Chua, remains debated~\cite{Ventra.2018}. 

Similarly, other nonlinear devices with memory such as memcapacitors and meminductors have been proposed~\cite{DiVentraIEEE2009,YinIEEE2015}, where the main difference is related to the input-output relation. Memristors relate voltage and current, memcapacitors relate voltage and charge and meminductors relate flux and current. The different input-output relations also provide different coupling mechanisms as well as different high frequency behaviour. Recently, memory devices have been studied as fundamental elements for neuromorphic computing~\cite{Park.2022,Borghetti2010,Wang2018NM, Li2018JPD}, offering potential for robust neuromorphic architectures~\cite{NatComm.2018, Song.2019, Gao.2020, Roldan.2022} beyond  von Neumann's architectures~\cite{Sebastian2020NatNanotechnol, KimiScience, Kundu2022VLSI}. 

On the other hand, quantum computing has shown the potential to revolutionize computer science with the first claims of quantum advantage, but always in the context of von Neumann's architecture. In this context, it is natural to think of quantum memory devices such as quantum memcapacitors, that is, memcapacitors working in the quantum regime. In recent years, quantum devices with memory properties have been proposed in platforms like superconducting circuits~\cite{Pfeiffer2016SciRep, PRapp.2016, PRapp.2018,Phys.Rev.Applied.2014, Sci.Rep.2016} and photonics~\cite{sanz2018, materials.2020}, with an experimental realization in 2021~\cite{Spagnolo.2021}. These proposals align with the emergence of neuromorphic quantum computing, which aims to develop quantum hardware and software implementations with brain-inspired devices~\cite{Markovica2020, Nakajima.2017, Fujii.2021}. Scalable quantum memcapacitors may also enable the development of analog devices that simulate brain-inspired functions, nonlinear models of materials, biology, and finance. Ongoing studies on coupled quantum memory devices have shown the nontrivial presence of quantum correlations in a memristive dynamics, a useful resource for interconnected quantum memristor arrays~\cite{Shubham.PRA, Kumar.2022}, as is suggested in reservoir computing paradigm~\cite{Spagnolo.2021}.

In this work, we propose a superconducting circuit design for the feasible implementation of a microwave quantum memcapacitor and its extension to multipartite arrays. Our proposal employs two coupled LC oscillators and a SQUID to adjust the effective frequency of one of the oscillator through an external magnetic flux. Such external magnetic flux depends on a weak measurement over the other oscillator, thus implementing a feedback process. We characterize the memcapacitive response of the proposed device to the external voltage applied over one of the oscillators. To do this, we consider different separable and entangled initial states. Additionally, we explore the response of coupled devices, computing the quantum correlations during the memcapacitive dynamics, revealing a nontrivial behavior.

\section{Results and Discussion}

\subsection{Classical and quantum memory devices}
\label{Sec.2}
We characterize a memory system by its input-output relation \cite{IEEE.1976}
\begin{equation}
\label{Eq.1}
y (t)  = g\big[ x , u , t\big] u (t),
\end{equation}
where $y(t)$ and $u(t)$ stand for the output and input signal of the system, respectively, and are related through the response function $g[x,u,t]$. The response function also depends on a state variable $x$ whose dynamics is described by the equation
\begin{equation}
\label{Eq.2}
\dot{x}  = f\big[ x , u , t\big].
\end{equation}
In the context of electrical circuits, and specifically for memristors, the response function $g[x,u,t]$ is usually called memristance~\cite{IEEE.1976, Pershin.2013} and can be derived using linear response theory developed in 1957 by R. Kubo \cite{Kubo.1957}. We note that Eq. (\ref{Eq.1}) ensure pinched hysteresis curves since when the input becomes zero, the output also becomes zero. This feature extends directly to the quantum case for quantum memory devices in an ideal case, nevertheless issue can be relaxed according to the Kubo's response theory. Classical devices have restrictions on $f$ and $g$ to ensure passivity, which leads to requiring $f$ to be always positive. Such property do not extend to quantum memory devices in general as we will explain latter.

Now, we can define a quantum memory device in a similar way. We can consider observables $\langle \hat{y} (t)\rangle$ and $\langle \hat{u}(t) \rangle$ following a similar relation as Eq.~(\ref{Eq.1}) and Eq.~(\ref{Eq.2})
\begin{subequations}
	\begin{eqnarray}
		\label{Eq.3(a)}
		\langle \hat{y} (t)\rangle &=& G\big[\langle \hat{x} \rangle , \langle \hat{u} \rangle , t\big] \langle \hat{u} (t)\rangle,\\
		\label{Eq.3(b)}
		\langle \dot{\hat{x}} \rangle &=& F\big[ \langle \hat{x} \rangle , \langle \hat{u} \rangle , t\big].
	\end{eqnarray}
\end{subequations}
Here, $G\big[\langle \hat{x} \rangle , \langle \hat{u} \rangle, t\big]$ and $F\big[ \langle \hat{x} \rangle , \langle \hat{u} \rangle, t\big]$ are the quantum analog to the response and state variable function, respectively. Also, we note that this is an input-output  relation between expectation values of physical observables, the dynamics of which is described by quantum mechanical laws.

\begin{figure}
	\centering
	\includegraphics[width=.5\linewidth]{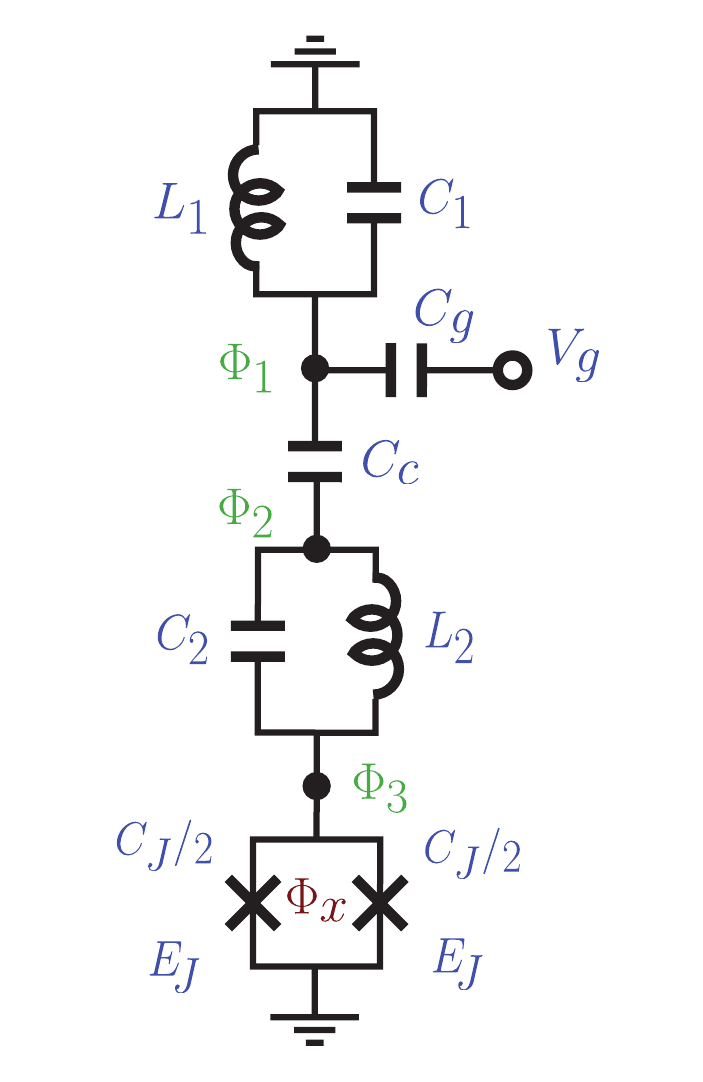}
	\caption{\textbf{Circuit design for the proposed memcapacitive device  in the microwave regime.} The input signal is given by $V_g$, and the feedback is provided by the magnetic flux $\Phi_x$ through the SQUID.}
	\label{Fig:circuit_model}
\end{figure}

Even though there are general methods to quantize electrical circuit elements with classical counterparts such as capacitors and inductors, or pure quantum ones such as Josephson junction~\cite{VoolIJCTA2017}, even complex elements as $n$-port nonreciprocal ones~\cite{ParraRodriguezPhysRevB2019,EgusquizaPhysRevB2022}, to the best of our knowledge there is no general formulation for the quantization of memory devices, despite efforts made on other platforms~\cite{Pfeiffer2016SciRep,PRapp.2016,PRapp.2018,Phys.Rev.Applied.2014,Sci.Rep.2016,sanz2018,materials.2020,Spagnolo.2021}. For this reason the characterization of quantum devices with memory properties takes relevance. In this context our proposal for a quantum memcapacitor in the microwave regime aims to develop and characterize scalable quantum components for neuromorphic quantum devices.

It is important to mention that in classical memory devices such as memcapacitors, memristors and meminductors, passivity is an essential property. Nevertheless, in quantum technologies this condition has been relaxed, and only the nonlinear response with memory signatures is studied. This is the case of the experimental quantum memory device reported in Ref. \cite{Spagnolo.2021}, and other theoretical proposals.

\subsection{The model}
\label{Sec.3}
We consider the circuit shown in Fig. \ref{Fig:circuit_model}, which is composed of two LC oscillators, each with a capacitance $C_j$ and inductance $L_j$, and coupled by a capacitor $C_c$. One of the resonators, labeled with $j=2$, is coupled galvanically to a SQUID, which consists of a closed loop with two Josephson junctions. The SQUID in the circuit acts as a Josephson junction with capacitance $C_J$ and tunable Josephson energy, given by $2E_{J}|\cos(2\pi\Phi_{x}/\Phi_{0})|$, which depends on the external magnetic flux $\Phi_x$ threading the SQUID. Here $\Phi_{0}=h/(2e)$ is the superconducting magnetic flux quantum. The role of the SQUID in this design is to change the cavity properties using the external magnetic flux $\Phi_x$ as feedback. Finally, we provide an input signal using a voltage source coupled capacitively, as is shown in Fig.~\ref{Fig:circuit_model}. The Lagrangian that describes our circuit reads
\begin{equation}
\label{Eq.4}
	\begin{split}
		\mathcal{L}= &\frac{ C_1}{2} \dot{\Phi} _1^2-\frac{\Phi_1^2}{2L_1}+\frac{C_g}{2} (\dot{\Phi} _1-V_g)^2
		+\frac{C_c }{2} (\dot{\Phi} _1-\dot{\Phi} _2)^2\\& +\frac{C_2}{2}(\dot{\Phi} _2-\dot{\Phi} _3)^2-\frac{(\Phi_2-\Phi_3)^2}{2L_2}
		+\frac{C_J}{2}\dot{\Phi} _3^2\\&
		+2E_J\cos(\varphi_x)\cos(\varphi_3),
	\end{split}
\end{equation}
where $\varphi_3 = 2\pi\Phi_3/\Phi_0$ is the superconducting phase and  $\varphi_x = 2\pi\Phi_x/\Phi_0$. Using the Legendre transformation and second-quantization techniques, we obtain the system Hamiltonian $\mathcal{\hat{H}}$ as (for detailed derivation, see Methods in section~\ref{AppA})
\begin{eqnarray}
	\label{Eq.5}
	\mathcal{\hat{H}}(\Phi_{x}) =\sum_{\ell=1,2}\bigg[\omega_\ell(\Phi_{x})\hat{a}^{\dagger}_\ell\hat{a}_\ell+iG_{g_{\ell}}(\Phi_{x},V_g)(\hat{a}_\ell^{\dagger}-\hat{a}_\ell)\bigg]\nonumber \\
	+\lambda^{+}(\Phi_{x})(\hat{a}_1^{\dagger}\hat{a}_2+\hat{a}_1\hat{a}_2^{\dagger})+\lambda^{-}(\Phi_{x})(\hat{a}_1^{\dagger}\hat{a}_2^{\dagger}+\hat{a}_1\hat{a}_2),
\end{eqnarray}
where we adopt the convention $\hbar=1$. Here, $\omega_\ell(\Phi_{x})$ is the effective frequency of the $\ell$th resonator, modified by the external magnetic signal in the SQUID. The effect of the voltage source $V_g$ over each cavity is represented by $G_{g_{\ell}}(\Phi_{x}, V_g)$, which also depends on the external flux $\Phi_x$. The effective coupling strength between resonators, $\lambda^{\pm}(\Phi_{x})=I_{12}(\Phi_{x})\pm G_{12}(\Phi_{x})$, comprises both an inductive contribution, $I_{12}(\Phi_{x})$, and a capacitive contribution, $G_{12}(\Phi_{x})$. It is important to note that all coefficients in the Hamiltonian depend on the external flux $\Phi_{x}$. We remark that we have used the high-plasma frequency and low-impedance approximation~\cite{Nori.2010, Delsing.2011}, which enables us to express $Q_{3}$ and $\Phi_{3}$ in terms of the other two charge and flux variables.

Therefore, our proposed device consists of two coupled harmonic oscillators connected via a capacitor. The effective frequency of each resonator is time-dependent and a function of the external magnetic flux through the SQUID. We update this flux using a feedback mechanism based on weak measurements applied to one of the oscillators. As a result, this process introduces changes in the response of the quantum devices, leading to effective memory properties.

To analyze the memory behavior of our device, we consider as input signal the voltage $V_g$, and the output signal as the signal in the node $\Phi_2$. We study the response of the charge operator in the oscillator 2, considering quantum feedback through the SQUID based on weak measurements on the oscillator 1. We call the initial state of the total system as $\ket{\Psi(0)}$. We update the external magnetic flux according to a cosine function, remember that trigonometric function has been successfully used previously, which for short time can be approximated as
\begin{figure*}[htp]
	\centering
	\includegraphics[width=0.9\linewidth]{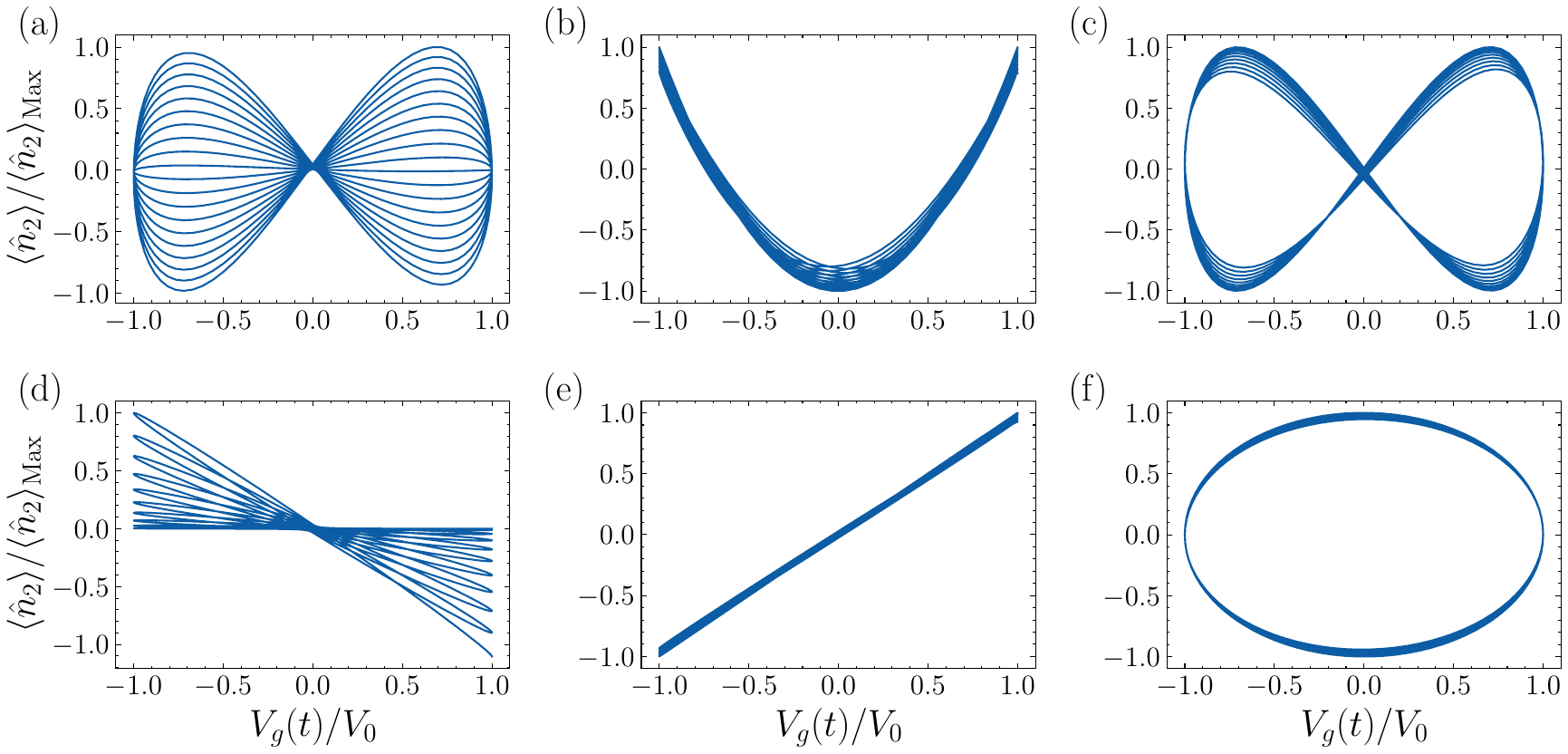}
	\caption{\textbf{Dynamic response of microwave quantum device under the action of the input voltage for different initial states.} (a) $\ket{\Psi(0)}=\ket{0}\otimes\ket{0}$, $\omega_{\nu}=\pi/5.92~\omega_{1}$ and (b)-(c) $\ket{\Psi(0)}=\ket{0}\otimes\ket{\Psi(\eta,\chi)}$ with $\eta=\pi/2$, for (b) $\chi=0$, for (c) $\chi=\pi/2$, and $\omega_{\nu}=\pi/5.94~\omega_{1}$. (d)-(f) High-frequency response of the aforementioned cases at $2\omega_{\nu}$. The calculations used $V_{0} = 0.01 \mu V$. Coupling strengths $\mathcal{G}_{12}/\omega_1 =5.294\times 10^{-3}$, $\mathcal{I}_{12}/\omega_1 = 1.998\times10^{-4} $.}
	\label{Fig:response_vacuum}
\end{figure*}
\begin{equation}
			\label{Eq.6}
			\frac{\Phi_{x}^{(j)}}{\Phi_{0}}=c_1-c_2\langle \hat{\varphi}_1(t_j) \rangle^2.
\end{equation}
where $c_1$ and $c_2$ are constants and $t_{j}=j\Delta t$. During the time windows $[t_j, t_{j+1}]$, the magnetic flux is constant and given by $\Phi_x^{(j)}$. To ensure continuous feedback, we use the condition $\omega{\Delta t} \ll 1$ where $\omega$ is the input voltage frequency. This means the time window $\Delta t$ is much smaller than the input voltage oscillation period, allowing fast updates. Experimentally, we can also replace each LC oscillator with a coplanar waveguide resonator, considering only the fundamental mode, where weak measurements can be performed over microwave photons, and the measurement outcome can be used to provide analog quantum feedback \cite{Irfan.2012, DiCarlo.2014, Lloyd.2000}.

It is important to mention that in order to get the input-output relation as Eq. (\ref{Eq.3(a)}) and Eq. (\ref{Eq.3(b)}), it is necessary to integrate the Schr\"{o}dinger equation in the Heisenberg picture, that is, $\dot{O}=(i/\hbar)[H,O]$ for the different operators $O$. This is a challenging task due to the large dimension of the involved systems, as well as the nontrivial relation between observables introduced by the feedback process. An alternative approach is to obtain the Krauss representation of the dynamics, which directly provides the input-output relation. Nevertheless, to obtain the analytical form of the Krauss representation is also quite involve for most quantum systems, being an active research area. 

Nonetheless, we can have an intuitive understanding of the memory relations in our system. The internal variable in the circuit should be a function of the magnetic flux $\Phi_x$ which depends on the output of a weak measurement over the dynamical variable $\hat{\varphi}_1$, that carries information of the input signal $V_g$. As the properties of the device depend on $\Phi_x$, the rate of change in the value $\Phi_x$, will depend on its instantaneous value, being natural to think in a relation of the form given by Eq. (\ref{Eq.3(b)}). In a similar way, the output signal must be dependent on the internal variable $\Phi_x$, since this magnetic flux changes the physical properties of the system. Additionally, the output variable depends on the input signal via resonance condition, which changes in time due to the dependence of the Hamiltonian (\ref{Eq.5}) with the magnetix flux. Thus,  Eq. (\ref{Eq.3(a)}) becomes a natural ansatz for the input-output relation. The input-output relation of the memory device is given by the external voltage $V_{g}(t)$ and  $\langle\hat{n}_{2}(t)\rangle$. In what follows, we will analyze the response of the observable $\langle\hat{n}_{2}(t)\rangle$ for different initial states and changing the frequency of the external voltage with $V_{g}(t)=V_{0}\cos(\omega_{\nu}t)$.

We remark that we will consider a driving frequency of the input signal of the order of the frequency of resonator 1, $\omega_{\nu} \sim \omega_{1}$. This means that timescale of a cycle of the input signal is much smaller than the relaxation time of the system, $\tau{s}$, which can be taken as $\tau{s} \sim 10^{3}\omega_1^{-1}$ \cite{Ridolfo.2012, Yoshioka.2023}. Therefore, we can analyze the time evolution of our system for tens of cycles of the input signal without considering the interaction with the environment.

\subsection{Single microwave quantum memcapacitor}
\label{Sec.4}
In this section, we study the response of the single device to different initial states, and driving frequencies. We remark that the values of $c_1$ and $c_2$ in Eq. (\ref{Eq.6}) and the driving frequency $\omega_{\nu}$ have been obtained by numerical optimization to maximize the memory properties for the different cases considered. First, we consider non-correlated initial states. Specifically, we consider the first resonator in the vacuum state and the second in superposition between zero and one photon, coherent and squeezed states, respectively. Afterwards, we consider correlated initial states for both oscillators, such as Bell-like, NOON, and cat states. The values of the constants in Eq.~ (\ref{Eq.6}) are considered $c_1 = 1.84$ and $c_2 = 0.08$ throughout the analysis. The circuit parameters used in our calculations are summarized in Table \ref{tab:Table1} in Methods, section \ref{A.2}.

We study the response of the device modifying the input voltage frequency for two particular values, where we expect to observe pinched hysteresis at a high-frequency regime, where the dynamics tend to behave as a linear resistor. These two behaviors are the fingerprints of a memcapacitor system~\cite{DiVentraIEEE2009}.

\subsection{Non-correlated inputs}
\begin{figure*}[t!]
	\centering
	\includegraphics[width=0.9\linewidth]{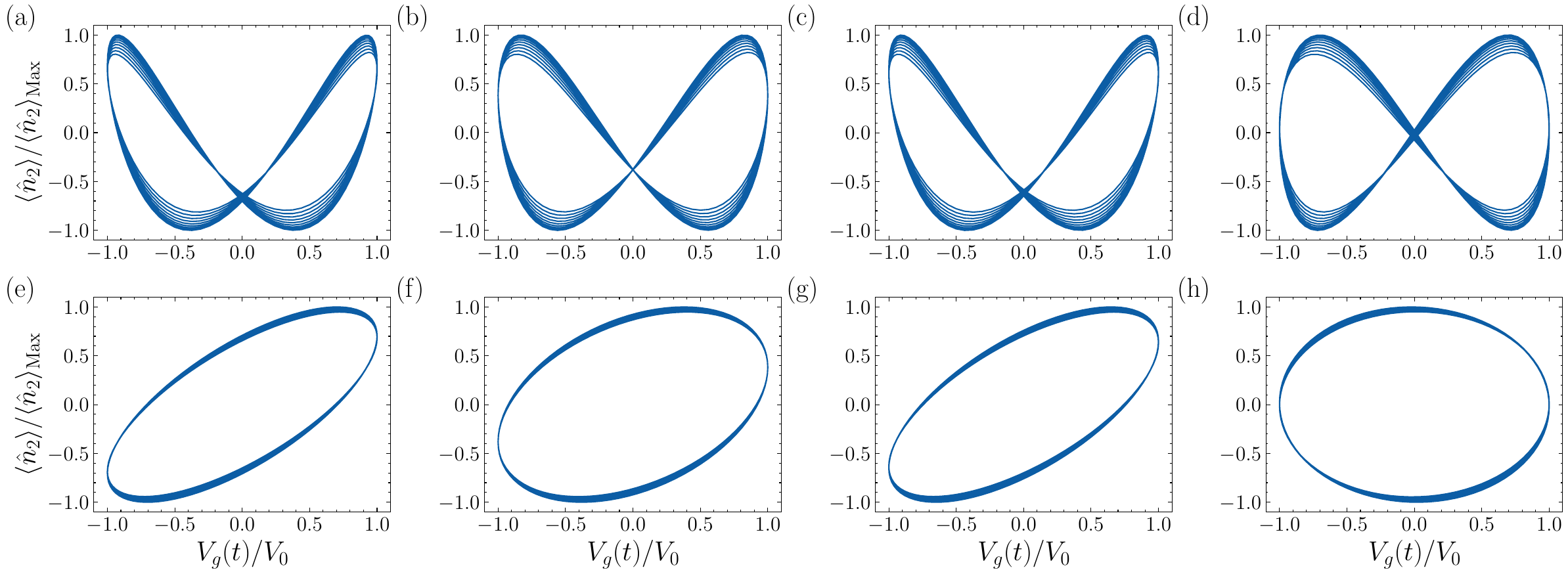}
	\caption{\textbf{Dynamic response of the microwave quantum device under the action of the input voltage for different initial states.} (a)-(b) $\ket{\Psi(0)}=\ket{0}\otimes\ket{\alpha}\equiv\ket{0}\otimes\ket{re^{i\varphi}}$ with $r=\pi/4$, (a) $\varphi=\pi/4$,  whereas (b) $\varphi=\pi/8$. (c)-(d)  $\ket{\Psi(0)}=\ket{0}\otimes\ket{\Psi(\alpha,\xi)}\equiv\ket{0}\otimes\hat{S}(Re^{i\theta})\ket{re^{i\varphi}}$, with (c) $R=0.1$, $\theta=\pi/4$ and (d) $R=1$, $\theta=\pi/4$. (e)-(h) High-frequency response ($2\omega_{\nu}$) for the previous states. For both cases, the driving frequencies are identical $\omega_\nu=\pi/5.92\omega_1$. Coupling strength $\mathcal{G}_{12}/\omega_1 =5.294\times 10^{-3}$, $\mathcal{I}_{12}/\omega_1 = 2.00\times10^{-4} $.}
		\label{Fig:response_coherent1}
\end{figure*}

We start our analysis considering initial vacuum states for both resonators $\ket{\Psi(0)}=\ket{0}\otimes\ket{0}$. Figure~\ref{Fig:response_vacuum}{(a)} shows the evolution of the charge $\langle \hat{n}_2\rangle$ as a function of the external voltage $V_g$. For the input voltage frequency $\omega_{\nu} = \pi/5.94\omega_1$, we can see that the curve is pinched at the origin, which can be considered as characteristic of a memcapacitor device~\cite {DiVentraIEEE2009}. On the other hand, for the input voltage frequency $2\omega_{\nu}$, high-frequency regime, we observe in Fig.~\ref{Fig:response_vacuum}(d) that the response tends to a line, which is another feature of memory behavior. It means our system, starting from the ground state, can be considered a memory device.
\begin{figure*}[t!]
	\centering
	\includegraphics[width=0.9 \linewidth]{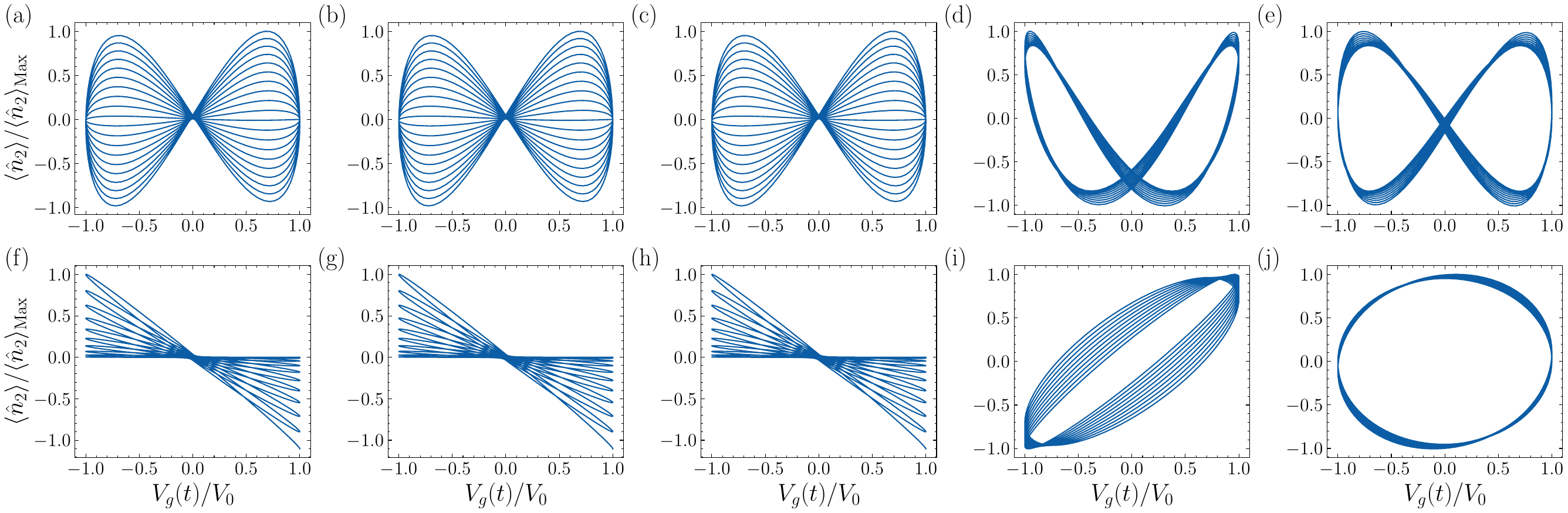}
	\caption{\textbf{Dynamic response of the microwave quantum device under the action of the input voltage when the system is initialized in different states.} (a)-(b)the Bell state, $\ket{\Psi(0)}=\cos\theta\ket{0,0}+\sin\theta\ket{1,1}$ where (a) $\theta=\pi/4$, and (b) $\theta=\pi/16$, driving frequency $\omega_\nu=\pi/5.94\omega_1$. (c) Noon state, $\ket{\Psi(0)}=(\ket{2,0}+\ket{0,2})/\sqrt{2}$. Cat state  $\ket{\Psi(0)}=(\ket{\alpha,0}+\ket{0,\alpha})/\sqrt{2}$ with $\alpha=re^{i\varphi}$, where (d) $r=\pi/2$, $\varphi=\pi/4$ and (e) $r=\pi/2$, $\varphi=\pi/2$ where $\omega_\nu=\pi/5.9\omega_1$. (f)-(j) High-frequency response ($2\omega_{\nu}$) for the previous states. We consider $V_{0} = 0.01 \mu V$, whereas the coupling strengths are $\mathcal{G}_{12}/\omega_1 =5.294\times 10^{-3}$, $\mathcal{I}_{12}/\omega_1 = 1.998\times10^{-4} $.}
	\label{response_bell 1}
\end{figure*}

For the case of the initial state $\ket{\Psi(0)}=\ket{0}\otimes\ket{\psi(\eta,\chi)}$, with $\ket{\psi(\eta,\chi)} = \cos(\eta/2)\ket{0}+e^{i\chi}\sin(\eta/2)\ket{1}$. The dynamical response of the device will be modified as long as we change phase $\chi$. Here, we will consider states with the same amplitudes $\eta=\pi/2$ and choose two values for the relative phase $\chi=\{0,\pi/2\}$. Figure ~\ref{Fig:response_vacuum}(b)-(c) shows the dynamical response of the device for the two mentioned values for the phase $\chi$ at driving frequency  $\omega_{\nu} = \pi/5.92\omega_1$. We notice that the value of $\chi$ plays the role of control for the memory  feature of our device. For $\chi=0$, we do not observe a hysteresis loop. However, adjusting  $\chi=\pi/2$, it shows a stable hysteresis loop. Moreover, at high-driving frequency, for phase $\chi=0$, we observe a line, and for $\chi=\pi/2$, we get a circle, which means that the phase $\chi$ also modifies the memory effects for the high-frequency regime.

It is interesting to also consider classical initial states, along the manuscript we call classical states to states that saturate the uncertainty relation and that have a non negative Wigner function. Under this definition coherent and squeeze states, can be named classical. We consider coherent states for the second oscillator, $\ket{\Psi(0)}=\ket{0}\otimes\ket{\alpha}$. We characterize the coherent state by its amplitude and phase through the relation $\alpha = r e^{i\varphi}$. We consider $r=\pi/4$ and $\varphi=\{\pi/4,\pi/8\}$. Figures~\ref{Fig:response_coherent1}(a)-(b) show the dynamical response for the two different phases. In both cases, the expectation value of $\langle \hat{n}_{2}(t)\rangle$ all exhibit the pinched hysteresis curve. Notice that in this case, the phase does not considerably affect the memcapacitor behavior as in the previous case. Also, we observe that for the high-frequency in Fig.~\ref{Fig:response_coherent1}(e)-(f), we obtain curves with oscillatory features again.

Finally, we can consider another type of classical states like squeezed states defined as $\ket{\Psi(\alpha,\xi)}=\hat{S}(\xi)\ket{\alpha}$. Here, $\ket{\alpha}$ is the same coherent state as defined in the previous section, and $\hat{S}(\xi)=\exp(\xi \hat{a}^{2}-\xi^{*} \hat{a}^{\dag~2})$ is the squeeze operator. Here, $\xi = Re^{i\theta}$ is also a complex number that characterizes the amount of squeezing and over which resonator quadrature will be applied. Thus, the initial state of the system is given by $\ket{\Psi(0)}=\ket{0}\otimes\ket{\Psi(\alpha,\xi)}$. In Fig.~\ref{Fig:response_coherent1}(c)-(d), we observe the input-output dynamics for different squeezing parameters. Specifically we use for Fig.~\ref{Fig:response_coherent1}(c) and (d) $R=0.1$ and $R=1$, respectively, and $\theta=\pi/2$ in both cases. In the high-frequency regime of the input voltage, we see that the system response again looks like an oscillator. 

We note that for all classical states (coherent, squeeze, and vacuum) as initial state, our coupled device shows memory properties in each memcapacitor, making our proposal suitable as a memdevice, at least for classical initialization.

\subsection{Correlated input}
\label{sec.5}
An interesting feature of quantum mechanics is the emergence of quantum correlations, which are useful resources to approach quantum advantage in quantum computing. Then, it is important to calculate the dynamical response of our devices for correlated inputs. Specifically, in this section, we consider a Bell-like state, the NOON state, which is a generalization of Bell states for higher photon numbers, and finally, we consider an initial cat state. For the case of Bell-like states, we consider initial superpositions of the form $|\Psi (0) \rangle= \cos(\theta)|0,0\rangle +\sin(\theta)|1,1\rangle$. The dynamics of the system is shown in Fig.~\ref{response_bell 1}(a) for $\theta=\pi/4$ and Fig.~\ref{response_bell 1}(b) for $\theta=\pi/16$. These two figures show similar results with the vacuum state, which suggests that the memory properties captured by the hysteresis loop are insensitive to the amount of entanglement. Also, for the high-frequency regime, see Fig. \ref{response_bell 1}(f)-(g), the dynamics tends to a line as the vacuum case, being again insensitive to the initial entanglement in the device. For the case of a NOON state, where we consider $\ket{\Psi(0)} = (\ket{2,0}+\ket{0,2})/\sqrt{2}$, again the dynamics present the same shape as can be seen in Fig.~\ref{response_bell 1}(c) and Fig.~\ref{response_bell 1}(h) (high-frequency regime). Finally, for entangled coherent or cat states, we consider $|\Psi (0)\rangle= (|\alpha,0\rangle +|0,\alpha\rangle)/\sqrt{2}$ with $\alpha = r e^{i\varphi}$. Here, the dynamics is close to the coherent state case as shown in Fig. \ref{response_bell 1}(d) and (e) for the cases of $\varphi=\pi/4$ and $\varphi=\pi/2$. In both cases with $r=\pi/2$, as well as for the high-frequency regime, that is Figs.~\ref{response_bell 1}(i)-(j).  These numerical results prove that our proposal keeps the behavior for entangled initial states between both resonators, which means entanglement between the internal variable and the output.

\subsection{Coupled microwave quantum memcapacitors}

\label{ Sec.6}

\begin{figure}
	\centering
	\includegraphics[width=.7\linewidth]{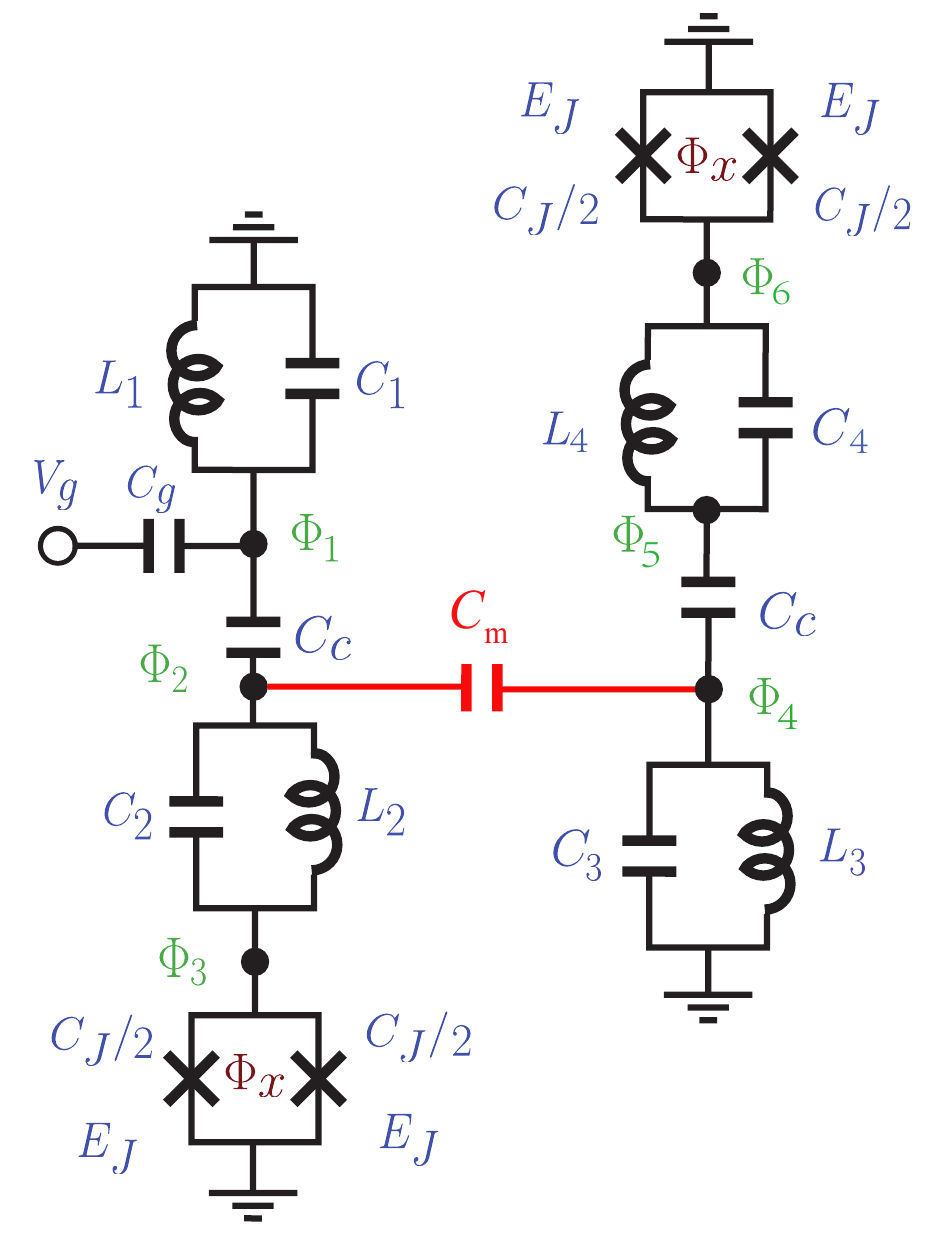}
	\caption{\textbf{Coupling of two microwave quantum memcapacitors}. The devices are coupled by a capacitor $C_m$ (red color). The second node, $\Phi_2$ in the first device, serves as the input signal to the second device.}
	\label{Fig:coupled circuit_model}
\end{figure}

We now consider a capacitive coupling between our proposed microwave quantum memcapacitor. Specifically, we consider a coupling between the input node of one device and the output node of the other device as is shown in Fig.~\ref{Fig:coupled circuit_model}. We have inverted the second microwave quantum memcapacitor to minimize the crosstalk effect between the SQUIDs. The circuit Hamiltonian for the coupled devices reads (see Methods section ~\ref{AppB} for the complete derivation)
\begin{widetext}
	\begin{eqnarray}\nonumber
		\label{Eq.7}
		\hat{\mathcal{H}}_{2M}&=&\sum_{\ell=\{1,2\}}\bigg[\omega_{\ell}(\Phi_{x})\hat{a}^{\dag}_{\ell}\hat{a}_{\ell}+iG_{g\ell}(\Phi_{x},t)(\hat{a}^{\dag}_{\ell}-\hat{a}_{\ell})+ \Omega_{\ell}(\Phi_{x})\hat{b}^{\dag}_{\ell}\hat{b}_{\ell} + iJ_{g\ell}(\Phi_{x},t)(\hat{b}^{\dag}_{\ell}-\hat{b}_{\ell})\bigg]\\\nonumber
		&+&\lambda^{+}(\Phi_{x})(\hat{a}_1^{\dagger}\hat{a}_2+\hat{a}_1\hat{a}_2^{\dagger})+\lambda^{-}(\Phi_{x})(\hat{a}_1^{\dagger}\hat{a}_2^{\dagger}+\hat{a}_1\hat{a}_2)+\Lambda^{+}(\Phi_{x})(\hat{b}_1^{\dagger}\hat{b}_2+\hat{b}_1\hat{b}_2^{\dagger})+\Lambda^{-}(\Phi_{x})(\hat{b}_1^{\dagger}\hat{b}_2^{\dagger}+\hat{b}_1\hat{b}_2)\\
		&+&\sum_{{j,k}=1}^{2}\bigg[\gamma^{+}_{j,k}(\Phi_{x})(\hat{a}_j^{\dagger}\hat{b}_k+\hat{a}_j\hat{b}_k^{\dagger})+\gamma^{-}_{j,k}(\Phi_{x})(\hat{a}_j^{\dagger}\hat{b}_k^{\dagger}+\hat{a}_j\hat{b}_k)\bigg].
	\end{eqnarray}
\end{widetext}
Here, $\hat{a}_{\ell}$ and $\hat{b}_{\ell}$ stand for the bosonic annihilation operators for each LC oscillator from the $\ell$th memcapacitive quantum device. Also, $\omega_{\ell}(\Phi_{x})$ and $\Omega_{\ell}(\Phi_{x})$ are the resonator frequency of each microwave quantum memcapacitor, while $G_{{g\ell }}(\Phi_{x})$ and $J_{g\ell }(\Phi_{x})$ correspond to the coupling strength between the resonators with the gate voltage. Moreover, $\lambda^{\pm}(\Phi_{x})$ and $\Lambda^{\pm}(\Phi_{x})$ are the coupling strength between the different nodes of each microwave quantum memcapacitor, whereas $\gamma_{j,k}^{\pm}(\Phi_{x})$ is the coupling strength between different devices. We will analyze the coupled case using the same initial state, non-correlated and correlated inputs, for each device of the previous section.

\subsection{Non-correlated inputs for microwave quantum memcapacitors}

\label{ Sec.7}

We analyze the dynamic response of the memcapacitive variable of the coupled device that corresponds to the second oscillator of each subsystem, labeled as oscillator 2 and oscillator 4. We study the evolution of a subsystem during the timescale $T=10(2\pi)/\omega_{\nu}$, with $\omega_{\nu}$ as the driving frequency of the input voltage. Notice that the coupling capacitance $C_{m}$ modifies the frequency of both microwave quantum memcapacitor, which leads to a slight change in the conditions required for the input voltage to achieve memory behavior in both devices. The parameter values used in the analysis are provided in Table \ref{tab:Table2} in Methods, section \ref{App.B2}.

\begin{figure}
	\centering
	\includegraphics[width=0.9\linewidth]{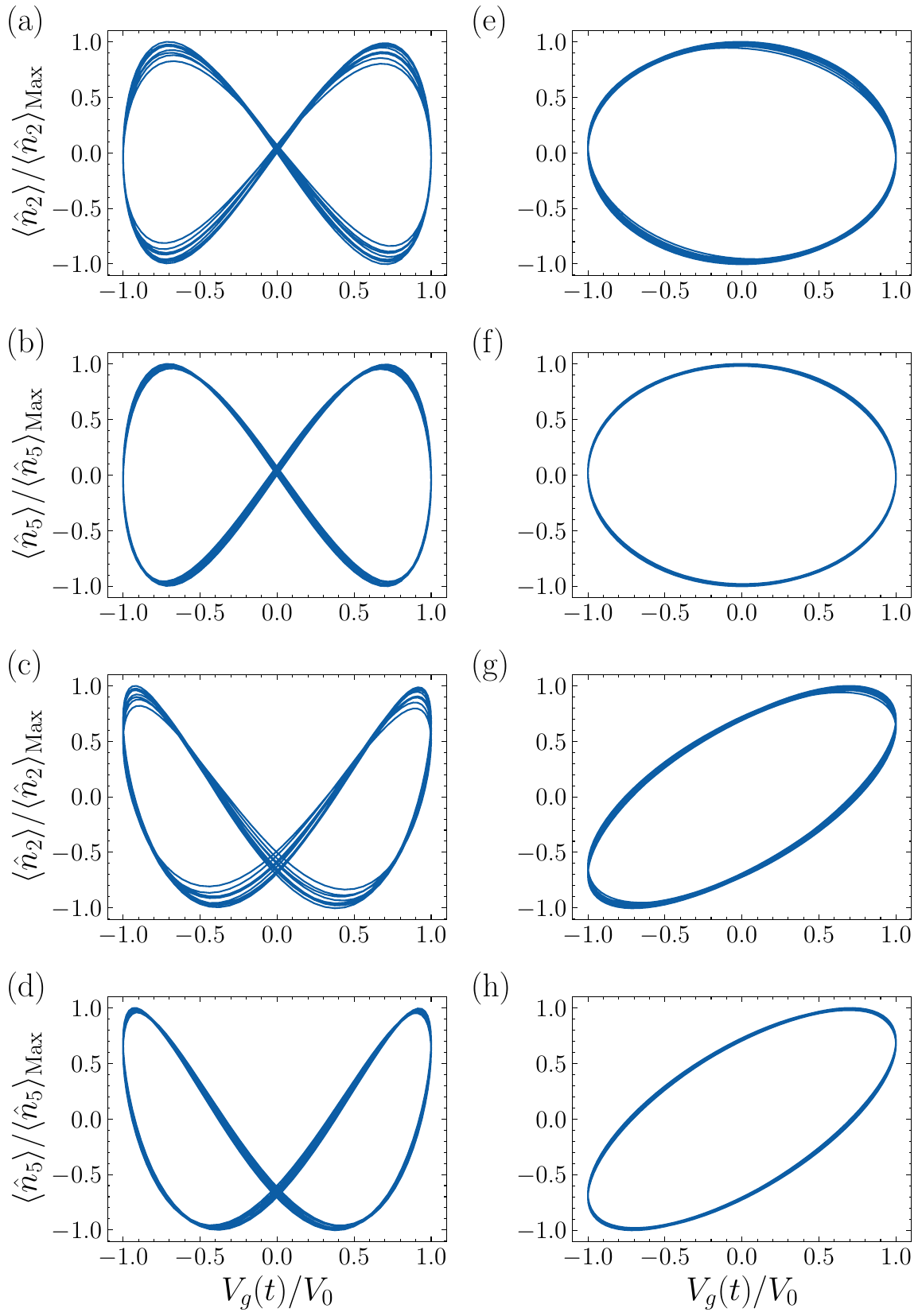}
	\caption{\textbf{Dynamic response of the coupled microwave quantum memcapacitive devices under the action of the input voltage for different states.} (a)-(b) superposition state, $\ket{\Psi (0)} = \ket{ 0,\psi(\eta,\chi)}\ket{0,\psi(\eta,\chi) }$, with $\eta=\pi/2$ and $\chi=0$. (c)-(d) Coherent state, $ \ket{\Psi (0)} = \ket{ 0,\alpha}\ket{0,\alpha}$, where $\alpha=r e^{i\varphi}$ with $r = \pi/2, \varphi = \pi/4$. (a)-(c) $\omega_\nu=\pi/6.33\omega_1$, (b) - (d) $\omega_\nu=\pi/6.25\omega_1$. (e)-(h) High-frequency response ($2\omega_{\nu}$) for the previous states.}
	\label{Fig:0.1superposition state2}
\end{figure}

We start our analysis by considering the initial state $\ket{\Psi (0)} = \ket{0,\psi(\pi/2,0)}\ket{0,\psi(\pi/2,0)}$, where $|\psi(\pi/2,0)\rangle = (\ket{0} +\ket{1})/\sqrt{2}$. We show the response with this initial condition in Fig.~\ref{Fig:0.1superposition state2}(a)-(b). We can observe that the observable $\langle \hat{n} \rangle$ exhibits memory behavior in both microwave quantum memcapacitors. At high frequency, shown in Fig.~\ref{Fig:0.1superposition state2}(e)-(f), we can observe that dynamics response form circumference, which corresponds to an oscillatory behavior.

Now, we consider coherent states of the form $\ket{\Psi (0)} = \ket{0,\alpha}\ket{0,\alpha}$, where $\alpha=(\pi/2) e^{i\pi/4}$. At voltage frequency $\omega_{\nu } = 0.5 \omega_1$ both devices show pinched hysteresis loop for $\langle \hat{n}\rangle$. Similar to the superposition state, we observe that the memcapacitive response of the second device is more stable than the first one (see Fig.~\ref{Fig:0.1superposition state2}(c)-(d)) where again the pinched curve does not shrink or expand. On the other hand, at high frequency ($2\omega_{\nu}$), the dynamical response of the variable $\langle\hat{n}\rangle$ corresponds again to an oscillator-like behavior.

\begin{figure}
	\centering
	\includegraphics[width=0.9\linewidth]{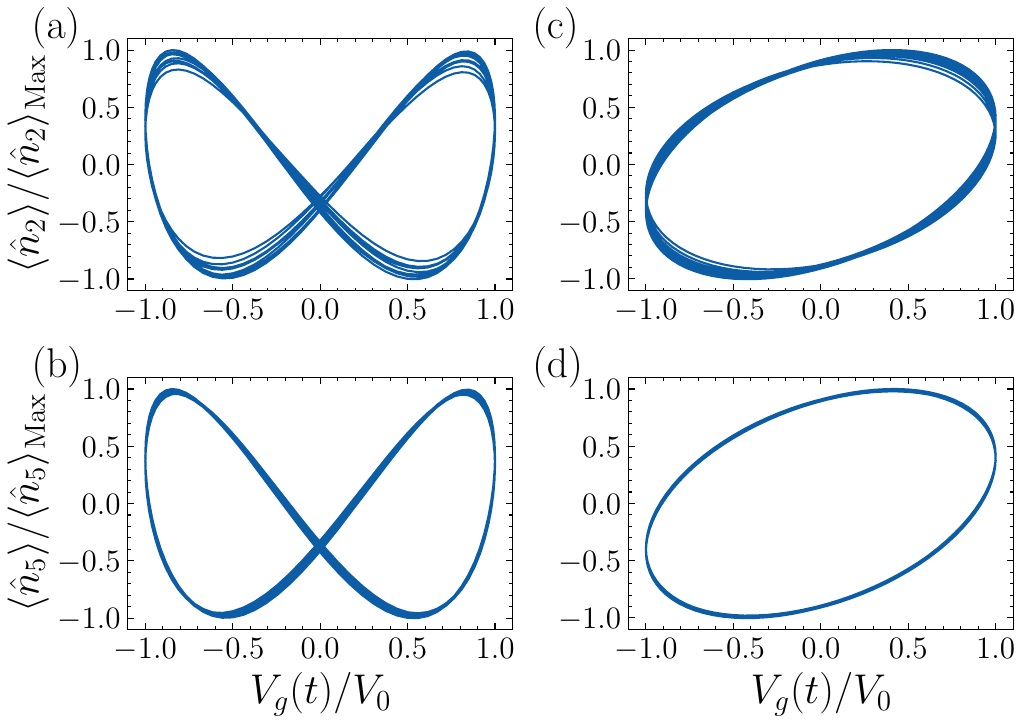}
	\caption{\textbf{Dynamic response of coupled microwave quantum devices utilizing squeezed states.} Initial state $\ket{\Psi (0)} = \ket{ 0,\alpha\xi}\ket{0,\alpha\xi }$, where $\alpha = r e^{i\varphi}$, $\xi = r e^{i\varphi}$ ($r = 0.1$, $\varphi = \pi/8$). (a) $\omega_\nu=\pi/6.33\omega_1$, and (b) $\omega_\nu=\pi/6.25\omega_1$. (c) and (d) are the corresponding high-frequency response of the device at double the frequency previously considered.}
	\label{Fig:0.1squeezed state1}
\end{figure}

Finally, we consider as the initial state the state $\ket{\Psi (0)} =\ket{0 ,\alpha\xi}\ket{0 ,\alpha\xi}$ corresponding to a squeezed state for the resonator 2 and 4, in this case we choose $\alpha=\xi=0.1e^{\pi/8}$. Similar to the uncoupled case, here we also observe memory behavior due to the pinched loop at voltage frequency $\omega_{\nu} = 0.5\omega_1$ (see Fig.~\ref{Fig:0.1squeezed state1}). The second memcapacitive system has a more stable response than the first one, maintaining unaltered its pinched hysteresis curve. For the high-frequency response, we see that the system exhibits an elliptical response, looking like an oscillator. This dynamics is similar to the single-device case using the squeezed states.

As a brief conclusion, we can observe that for the case of uncorrelated inputs, the memcapacitive behavior is preserved, obtaining curves with the same shape as in the uncoupled case. Also, we can see that the response of the microwave quantum memcapacitor that is further to the input voltage presents more stable dynamics. This can suggest that for a chain of microwave quantum memcapacitor with uncorrelated inputs, the dynamics will be more stable at the end of the chain.

\subsection{Correlated Input for microwave quantum memcapacitor}

\label{ sec.8}

As in the case of uncoupled devices, we can calculate the dynamics of coupled devices with initially correlated input states. We use similar states as in the case of a single device it means Bell, NOON, and cat states.

\begin{figure}
	\centering
	\includegraphics[width=0.9\linewidth]{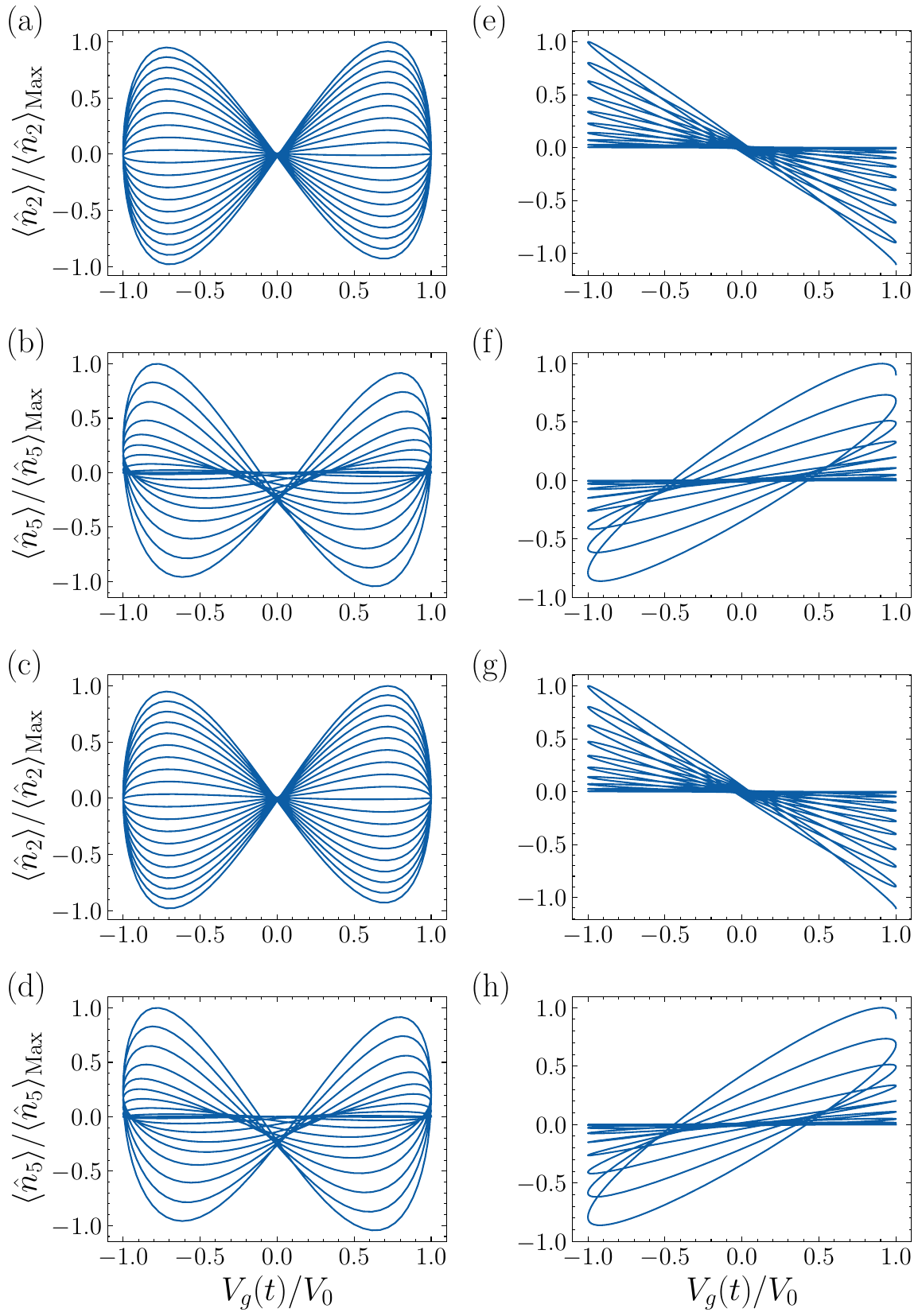}
	\caption{\textbf{Dynamic response of the coupled microwave quantum memcapacitive devices for different innitial states.} (a)-(b) Bell state $\ket{\Psi (0)} =  \ket{ \psi_B}\ket{\psi_B}$, with $|\psi_B\rangle= (\ket{0,0} +\ket{1,1})/\sqrt{2}$. (c)-(d) NOON state, $\ket{\Psi (0)} =\ket{ \psi_N}\ket{\psi_N }$, with $\ket{\psi_N}= (\ket{0,2} +\ket{2,0})/\sqrt{2}$.  The driving frequencies are (a)-(c) $\omega_\nu=\pi/6.3\omega_1$, and (b)-(d) $\omega_\nu=\pi/6.22\omega_1$. (e)-(h) High-frequency response ($2\omega_{\nu}$) for the previous states.} 
	\label{Fig:0.1bell state1}
\end{figure}

We start our analysis considering Bell state as the initial state, that is $\ket{\Psi (0)} =\ket{ \psi_B}\ket{\psi_B }$, where $\ket{\psi_B}= (\ket{0,0} +\ket{1,1})/\sqrt{2}$. Figures~\ref{Fig:0.1bell state1}(a)-(b) show the input-output dynamics for both memcapacitors for $\omega_{\nu} = 0.5\omega_1$. In this case, curves approach pinched hysteresis loops. We notice that the first device is more stable than the second one. For the high-frequency regime, we can note that the response of the observable $\langle\hat{n}_{\ell}\rangle$ squashes losing the memcapacitive properties  (see Fig.~\ref{Fig:0.1bell state1}(e)-(f)).

Next, we initialize the devices a in tensor product of NOON state of the form $\ket{\Psi (0)} = \ket{ \psi_N}\ket{\psi_N }$, where $\ket{\psi_N}= (\ket{0,2}+\ket{2,0})/{\sqrt{2}}$. Interestingly, we observe that for this initialization, the results are similar to Bell states, as depicted in Fig.~\ref{Fig:0.1bell state1}(c)-(d) for voltage frequency $\omega_{\nu} = 0.5\omega_1$. We need to mention that for Bell and NOON states, the expectation value in the number of photons is zero, which is related to the position of the pinched point in the curves, as pointed out in References~\cite{Shubham.PRA}. For NOON states, the high-frequency dynamics is similar to the Bell state as can be seen in Fig.~\ref{Fig:0.1bell state1}(g)-(h), where the devices tend to lose their memcapacitive properties.

\begin{figure}[htp]
	\centering
	\includegraphics[width=0.9\linewidth]{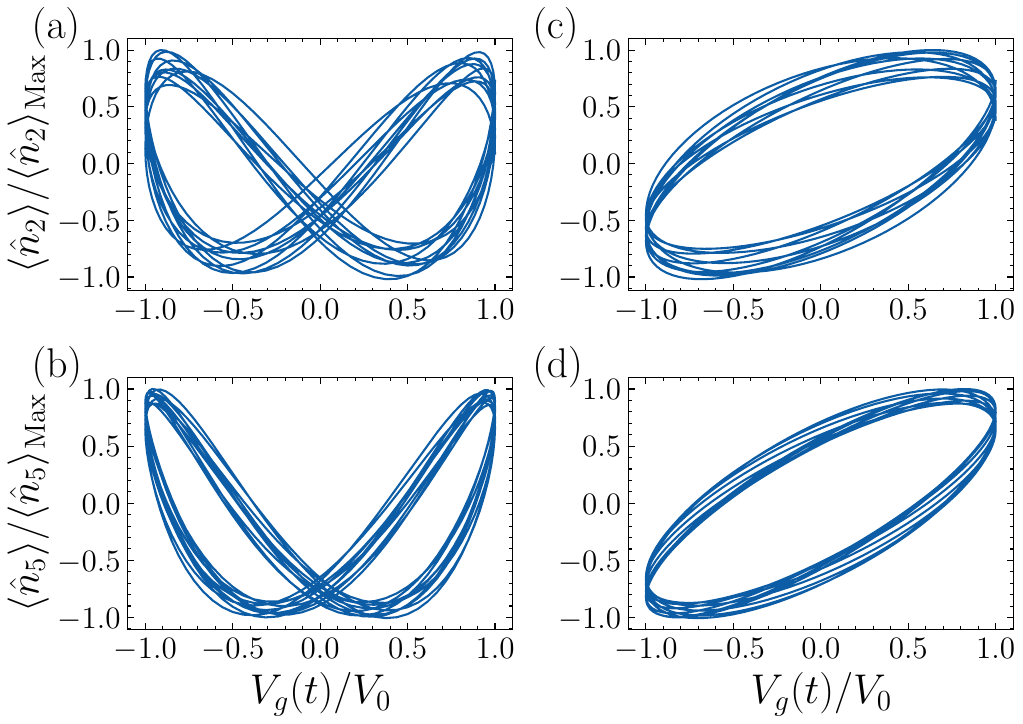}
	\caption{\textbf{Dynamic response of coupled microwave quantum devices using a product of entangled coherent state.} Initial state $\ket{\Psi (0)} = \ket{ \psi_C}\ket{\psi_C }$, where $\ket{\psi_C}=(\ket{\alpha, 0} +\ket{0,\alpha})/{\sqrt{2}} $ with $\alpha = re^{i\varphi}$ and $r = \pi/2$, $\varphi=\pi/4$. (a) The driving frequency $\omega_\nu=\pi/6.3\omega_1$, and (b) The driving frequency $\omega_\nu=\pi/6.22\omega_1$. Panels (c) and (d) are the high frequency response for the cases described in (a) and (b) respectivelly.} 
	\label{Fig:0.1cat entangled state1}
\end{figure}

Finally, we consider an initial cat state $\ket{\Psi(0)} = \ket{\psi_C}\ket{\psi_C} $, with $\ket{\psi_C}  = (\ket{\alpha, 0} +\ket{0,\alpha})/{\sqrt{2}}$. For input voltage frequency $\omega_\nu= 0.5\omega_1$ we can obtain pinched hysteresis curves, with a more stable response from the second device, as shown in Fig. \ref{Fig:0.1cat entangled state1}. On the other hand, at high frequency, the response in the coupled system produces a circle. Therefore, the memory dynamics is replaced by an oscillatory one. 

Again, all these results show that our proposal has memcapacitive quantum properties in each device when coupled in a suitable parameters regime, which can be switched with the frequency of the external input. It is important to mention that our proposal differs from the ideal memory device, where the input-output relation gives perfect close loops, nevertheless, as Kubo's response theory, these results can be linked with memory properties of the proposed device~\cite{DiVentraNanotechnology2013}.

\subsection{Quantum Correlations}
\label{Sec.9}

\begin{figure*}[htp]
	\centering
	\includegraphics[width=0.9\linewidth]{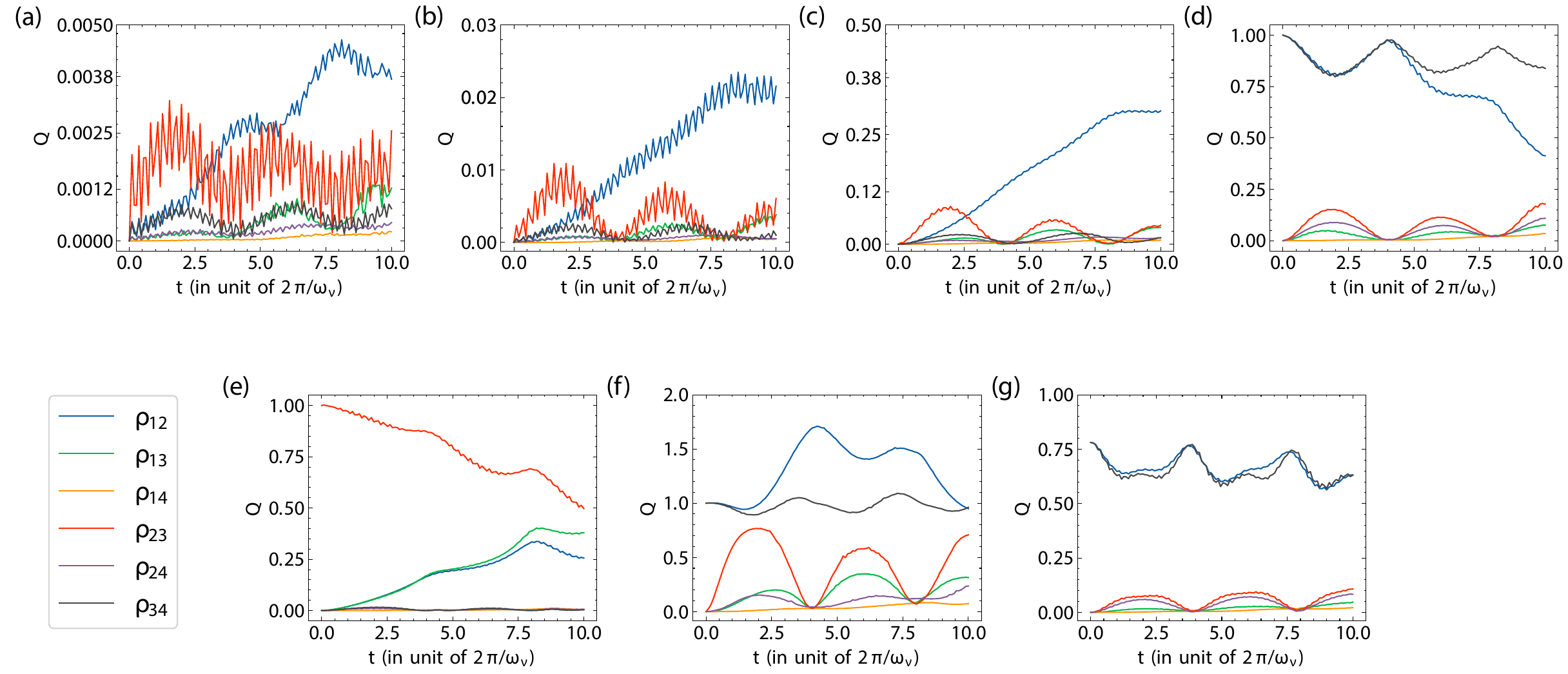}
	\caption{\textbf{Quantum discord between different resonators of the coupled QMs in different initial states.} $\rho_{ij}$ coresponds to the bipartite state of $i$th and $j$th resonator in the device. (a) Coherent state,$\ket{\Psi (0)} = \ket{0,\alpha}\ket{0,\alpha}$ with $ \alpha =\pi/2e^{i\pi/2} $. (b) Squeezed state, $\ket{\Psi (0)} = \ket{ 0,\alpha\xi}\ket{ 0,\alpha\xi}$, where $\alpha = r e^{i\varphi}$, $\xi = r e^{i\varphi}$ ($r = 0.1$, $\varphi = \pi/8$). (c) Superposition state, $ \ket{\Psi (0)} =  \ket{ 0,\psi(\pi/2,0)} \ket{ 0,\psi(\pi/2,0)}$, where $\ket{ \psi(\pi/2,0)} (\ket{ 0} +\ket{1})/\sqrt{2}$. (d) Bell state, $\ket{\Psi (0)} = \ket{\Psi_{B}}\ket{\Psi_{B}}$, (e) $\ket{\Psi (0)} = \ket{0}\ket{\Psi_{B}}\ket{0}$ where $|\psi_B\rangle  =\frac{1}{\sqrt{2}} (|00\rangle +|11\rangle)$. (f) Noon state, $\ket{\Psi (0)} =\ket{ \psi_N}\ket{\psi_N}$, where $|\psi_N\rangle= (|0,2\rangle +|2,0\rangle)/{\sqrt{2}}$. (g) Cat state, $\ket{\Psi (0)} = \ket{\psi_C}\ket{\psi_C} $, with $|\psi_C\rangle=(\ket{\alpha,0} +\ket{0,\alpha})/{\sqrt{2}} $.}
	\label{Fig:quantum discord}
\end{figure*}

We calculate the correlation embedded in the different resonators of our coupled device described by the reduced density matrix $\rho_{r_{i},r_{j}}=\textrm{Tr}_{r_{k},r_{l}}(\ketbra{\Psi}{\Psi})$, where we have traced out two of the resonators. As a measure of quantum correlations, we consider the quantum discord~\cite{Zurek.2002, Luo.PRA}, which considers all the correlations in a system that cannot be considered as classical correlations. Formally, quantum discord is defined as
\begin{eqnarray}
	\mathcal{Q}_{i,j} = S(\rho_{r_{i}})-\min_{\Pi_{r_{i}}^{m}}S(\rho_{r_{i},r_{j}|r_{j}}),
	\label{Eq.8}
\end{eqnarray}
where $S(\rho)=\rm{Tr}[\rho\log(\rho)]$ is the von Neumann entropy,  $\rho_{r_{i}}=\textrm{Tr}_{r_{j}}(\rho_{r_{i},r_{j}})$ and $\rho_{r_{i},r_{j}|r_{j}}$ is the density matrix after a projective measurement in the resonator $r_j$. The second term minimizes the von Neumann entropy for all the possible projective measurements in $r_j$. Such projective measurements can be written as $\Pi_{r_{j}}^{m}=\mathbb{I}_{r_{i}}\otimes U\ketbra{m}{m}U^{\dag}$. The $d$-dimensional unitary matrix $U$ can be written in terms of $d(d-1)/2$ two-level matrices as
\begin{equation}
	U=\prod_{k=1}^{d-1}\prod_{n=1}^{d-k}U_{k,n} .
	\label{Eq.9}
\end{equation}
Here, the matrix $U_{k,n}$ reads
\begin{equation}
\label{Eq.10}
	U_{k,n}=\left(\begin{array}{cccccccccc}
		1&0& ..& & & & && &\\
		0& 1& & & & & & & &\\
		..&&..&&&&&&\\
		&&&v_{k,k}&&&v_{k,k+n}&&&\\
		&&&&..&&&&&\\
		&&&&&..&&&&\\
		&&&v_{k+n,k}&&&v_{k+n,k+n}&&&\\
		&&&&&&&..&&..\\
		&&&&&&&&1&0\\
		&&&&&&&..&0&1\\
	\end{array}\right) ,
\end{equation}
where each matrix $U_{k,n}$ can be parametrized in terms of three angles, that is $v_{k,k}=\sin ({\phi_{1}})e^{i\phi_{2}}$, $v_{k+n,k}=\cos (\phi_{1})e^{i\phi_{3}}$, $v_{k,k+n}=\cos (\phi_{1})e^{-i\phi_{3}}$ and $v_{k+n,k+n}=-\sin ({\phi_{1}})e^{-i\phi_{2}}$. It means that the U in Eq.~(\ref{Eq.9}) can be parametrized by $3(d-1)d/2$ angles, which need to be optimized to minimize the second term in Eq.~(\ref{Eq.8}). To perform such an optimization process, we used the basinhopping algorithm from Ref.~\cite{basinhopping}.

Figure~\ref{Fig:quantum discord} shows the dynamics of the quantum correlation for the different resonators in the coupled microwave quantum memcapacitors configuration for different initial states. In Fig.~\ref{Fig:quantum discord} (a)-(c), we consider the bipartite state $\rho_{ij}$ for the $i$th and $j$th resonators, we consider initializations in the coherent state, squeezed-displaced state, and superposition state, respectively. As these initial product states possess no inherent correlations, the emergence of finite correlations over time indicates progressive quantum interaction between subsystems. These correlations experience oscillatory behavior, alternating between maximum and minimum values. We note that for the initial superposition state, we reach a non-negligible correlation between the microwave quantum memcapacitors. 

Next, we analyze the case when the system starts in an entangled state, Fig. \ref{Fig:quantum discord}(d)-(g), where we consider $\rho_{ij}$ to be in the bell, noon and cat states. In Fig.~\ref{Fig:quantum discord}(d), since the system is initialized in a maximally entangled state, $\rho_{12}$ and $\rho_{34}$ start with maximal quantum correlations that decay over time undergoing an oscillatory behavior. It is worth noticing that the oscillations produce a rise and decay in the correlations contained in $\rho_{12}$ and $\rho_{34}$ that coincide with the decay and rise of the correlations contained in $\rho_{13}$, $\rho_{14}$, $\rho_{23}$, $\rho_{24}$. This implies that the correlations between resonators of the individual microwave quantum memcapacitors get shared over time between the resonators of the different devices. This transfer of quantum correlations is a well-known phenomenon among multipartite systems~\cite{Solano.2008}, and their interplay with memory behavior has been reported recently in SQUID-based quantum memristor~\cite{Shubham.PRA, Kumar.2022}. 

Changing to a different initialization, the 2nd and 3rd resonators can be initialized in a Bell state. The plots corresponding to this configuration are shown in Fig.~\ref{Fig:quantum discord}(e). Due to this initial state configuration, the bipartite state $\rho_{23}$ starts from maximal correlations decaying with time and is accompanied by the increase in the correlations of the other bipartite states. We find a similar observation where the maximal (minimal) values of $\rho_{23}$ coincide with the minimal (maximal) values of the other states. Finally, we study the correlations when the system is initialized in the NOON and cat states, as is shown in Fig.~\ref{Fig:quantum discord}(f)-(g), where the transfer of correlations is also observed. Using the NOON state, the correlations evolve and go beyond unity since the number of photons in the resonators is 2, and therefore the maximal quantum correlation is not bounded to the unit. Similarly, using the cat state, the correlations depend on the value of $\alpha$ that determines the maximal correlations in the system. 

Our proposal can be easily implemented via capacitive coupling as is shown in Fig. \ref{Fig:coupled circuit_model}. This allows the implementation of more complex arrays of memcapacitive quantum devices, opening the door to the experimental implementation of neuromorphic quantum computing and simulation systems.

\section*{Conclusions}
\label{Sec04}
We proposed an experimentally feasible quantum memcapacitor device, a microwave quantum memcapacitor, using superconducting circuits in the microwave regime. Our design consists of two coupled resonators grounded through a SQUID, where one of the resonators plays the role of the main system and the other of the auxiliary feedback system. We observe memcapacitive quantum dynamics in pinched hysteresis curves in the expectation values of the charge variable when we introduce feedback through a magnetic flux in the SQUID. Such magnetic flux depends on weak measurements over the auxiliary resonator. We test these memcapacitive quantum behaviors for different initial states, from classical to entangled inputs.

We showed that our proposal can be easily extended to coupled microwave quantum memcapacitor, allowing for complex networks suitable for developing neuromorphic quantum computers and simulators. In this context, we proved that the memory properties are preserved when we couple two microwave quantum memcapacitors with different classical or quantum initial states. Finally, we displayed that the quantum correlations measured by the quantum discord present nontrivial behaviors, which are fingerprints of the quantumness of our device.

Also, it is necessary to highlight that quantum memory devices such as in Ref. \cite{Spagnolo.2021} has shown applicability in pattern recognition in the context of reservoir computing, outperforming the classical counterpart. This advantages is obtained by introducing coherence and quantum correlations into the reservoir of quantum memdevices. In this sense, our design offers great connectivity and scalability as two or more quantum devices can be capacitively coupled while still retaining their memory properties. Furthermore, with our design, the quantum memcapacitor become correlated in time being a suitable candidate for reservoir computing.

\section*{Acknowledgements}
F.A.-A. acknowledge financial support from Agencia Nacional de Investigaci\'on y Desarrollo (ANID): Subvenci\'on a la instalaci\'on en la Academia No. SA77210018, Fondecyt Regular No, 1231172, and Financiamiento Basal para Centros Cient\'ificos y Tecnol\'ogicos de Excelencia AFB 220001. F. A. C. L. acknowledge financial support from the German Ministry for Education and Research, under QSolid, Grant No. 13N16149.

\newpage

\onecolumngrid

\vspace{3cm}

\section{Methods}

\subsection{Derivation of the circuit Hamiltonian for a single microwave quantum memcapacitors}

\label{AppA}

This section is devoted to deriving the quantized Hamiltonian of the architecture shown in Fig.~\ref{Fig:circuit_model}. We derive the classical Hamiltonian and obtain a simplified version of the Hamiltonian using genuine approximations. Finally, we will quantize the Hamiltonian by promoting the charge and flux coordinates to the quantum operators.

\subsubsection{Classical Hamiltonian for a single microwave quantum memcapacitor}

The Lagrangian of the circuit given in Fig.~(\ref{Fig:circuit_model}) is
\begin{equation}
	\mathcal{L}=\frac{ C_1}{2} \dot{\Phi} _1^2-\frac{\Phi_1^2}{2L_1}+\frac{C_g}{2} (\dot{\Phi} _1-V_g)^2
	+\frac{C_c }{2} (\dot{\Phi} _1-\dot{\Phi} _2)^2+\frac{C_2}{2}(\dot{\Phi} _2-\dot{\Phi} _3)^2-\frac{(\Phi_2-\Phi_3)^2}{2L_2}
	+\frac{C_J}{2}\dot{\Phi} _3^2+2E_J\cos(\varphi_x)\cos(\varphi_3),
	\label{A1}
\end{equation}
where $\varphi_3 = 2\pi\Phi_3/\Phi_0$ is the superconducting phase, with $\Phi_0=h/2e $ as the quantum flux where $h$ is the Planck's constant and $2e$ is the Cooper-pair electric charge. Moreover, $\varphi_x = 2\pi\Phi_x/\Phi_0$ is the external flux through the SQUID. We calculate the canonical conjugate momenta (node charge) through the relation $ \mathcal{Q}_n=\partial \mathcal{L}/{\partial \dot{\Phi}_n}$,
\begin{multline}
	\mathcal{Q}_1 = (C_g+C_1+C_c)\dot{\Phi}_1-C_c\dot{\Phi}_2-C_gV_g, \; \; \;
	\mathcal{Q}_2 =-C_c\dot{\Phi}_1+C_c\dot{\Phi}_2+C_2(\dot{\Phi}_2-\dot{\Phi}_3 ), \; \; \;
	\mathcal{Q}_3 = C_2(\dot{\Phi}_3-\dot{\Phi}_2)+C_J\dot{\Phi}_3.
	\label{A2}
\end{multline}
By defining $\tilde{\mathcal{Q}}_1=\mathcal{Q}_1+C_gV_g$, $\tilde{\mathcal{Q}}_2=\mathcal{Q}_2$, $\tilde{\mathcal{Q}}_3=\mathcal{Q}_3$, the set of equations given in Eq.~(\ref{A2}) can be written as $\vec{\tilde{\mathcal{Q}}} = \hat{C}\vec{\dot{\Phi}}$. Here, $\vec{\tilde{\mathcal{Q}}}$ and $\vec{\dot{\Phi}}$ correspond to the charge and time derivative flux vector, respectively, and $\hat{C}$ is the capacitance matrix. Applying the Legendre transformation $\mathcal{H}= \sum_i\tilde{\mathcal{Q}}_i\dot{\Phi}_i-\mathcal{L}$, we obtain the circuit Hamiltonian as
\begin{equation}
	\begin{split}
		\mathcal{H} &=\frac{C^{-1}_{11}\tilde{\mathcal{Q}}_1^2}{2} +\frac{C^{-1}_{22}\tilde{\mathcal{Q}}_2^2}{2}+\frac{C^{-1}_{33}\tilde{\mathcal{Q}}_3^2}{2}
		+C^{-1}_{12}\tilde{\mathcal{Q}}_1\tilde{\mathcal{Q}}_2 +C^{-1}_{13}\tilde{\mathcal{Q}}_1\tilde{\mathcal{Q}}_3
		+C^{-1}_{23}\tilde{\mathcal{Q}}_2\tilde{\mathcal{Q}}_3+\frac{\Phi_1^2}{2L_1}+\frac{(\Phi_2-\Phi_3)^2}{2L_2}-2E_J\cos(\varphi_x)\cos(\varphi_3)
	\end{split} .
	\label{A5}
\end{equation}
Here, $C_{ij}^{-1}$ corresponds to the matrix elements of the inverse of the capacitance matrix given by
\begin{eqnarray}
	\nonumber
	\hat{C}^{-1}=\frac{1}{C^{\star}} \begin{pmatrix}
		C_c C_J+C_2 (C_c+C_J) & C_c (C_J+C_2) & C_2 C_c \\
		C_c (C_J+C_2) & ( C_c+C_1+C_g) (C_J+C_2) & C_2 ( C_c+C_1+C_g) \\
		C_2 C_c & C_2(C_c+C_1+C_g) & C_c C_g+C_2(C_c+C_g)+C_1 (C_c+C_2)
	\end{pmatrix},
	\label{A4}
\end{eqnarray}
where $C^{\star}=C_2C_c(C_1+C_g)+C_J(C_1(C_2+C_c)+C_cC_g+C_2(C_c+C_g))$. Notice that the system dynamics of our circuit depends on three degrees of freedom $\{\Phi_{1},\Phi_{2},\Phi_{3}\}$ corresponding to the two resonators and the SQUID, respectively. We may reduce the system dynamics in terms of $\{\Phi_{1},\Phi_{2}\}$ considering the high-plasma frequency ($\dot{\Phi}_3\ll\dot{\Phi}_{1(2)}$ ($\ddot{\Phi}_3\ll\ddot{\Phi}_{1(2)}$) and low-impedance regime (${\Phi}_3\ll{\Phi}_{1(2)}$) of the SQUID~\cite{Jing.2022}, obtaining the relation between the node charges as
\begin{equation}\label{A7}
	\tilde{\mathcal{Q}}_3= \bigg(\frac{-C_2 C_c}{C_2(C_1+C_g)+C_c(C_1+C_g+C_2)}\tilde{\mathcal{Q}}_1+\frac{-C_2(C_c+C_1+C_g)}{C_2(C_1+C_g)+C_c(C_1+C_g+C_2)}\tilde{\mathcal{Q}}_2\bigg).
\end{equation}
Next, we derive the relation between the node fluxes using the Euler-Lagrange equation $\partial \mathcal{L}/\partial\Phi_i - d(\partial\mathcal{L}/\partial \dot{\Phi}_i)/dt = 0$ and considering high-plasma frequency $\ddot{\Phi}_3\ll\ddot{\Phi}_{1(2)}$
\begin{multline}
	-\frac{\Phi_1}{L_1}+C_c\ddot{\Phi}_2-(C_c+C_1+C_g)\ddot{\Phi}_1=0, \; \; \;
	-\frac{\Phi_2}{L_2}-(C_c+C_2)\ddot{\Phi}_2+C_c\ddot{\Phi}_1=0, \; \; \;
	\frac{4\pi E_J\cos(\varphi_x)}{\Phi_0}\sin(\varphi_3)+C_2\ddot{\Phi}_2 =0.
	\label{A8}
\end{multline}
We get $\Phi_{3}$ from $\Phi_{1}$ and $\Phi_{2}$ using the second linearized regime of the Josephson junction~\cite{Blais.2021} i.e., $\sin(\varphi_3)=\varphi_3$ leading to
\begin{equation}
	\varphi_3=\frac{\Phi_0}{4\pi E_J\cos(\varphi_x)}\bigg(\alpha_1 \frac{\Phi_1}{L_1}+\alpha_2\frac{\Phi_2}{L_2}\bigg),
	\label{A10}
\end{equation}
where $\alpha_1 =\frac{C_2C_c}{(C_c+C_1+C_g)(C_c+C_2)-C_c^2} $ and $\alpha_2 =\frac{C_2(C_c+C_1+C_g)}{(C_c+C_1+C_g)(C_c+C_2)-C_c^2}$. We note that
\begin{equation}
\alpha_2 =\alpha_1 + \frac{C_2(C_1+C_g)}{(C_c+C_1+C_g)(C_c+C_2)-C_c^2}.
\end{equation}
To express the Hamiltonian in terms of Eq.~(\ref{A10}), we consider the low-impedance regime ($\Phi_3\ll\Phi_{1(2)}$), so that $\cos(\varphi_3)=1-\varphi_3^2/2$ and keep the potential energy of the SQUID up to the second order. Using Eq.~(\ref{A7}) and Eq.~(\ref{A10}) in Eq.~(\ref{A5}) we arrive at the Hamiltonian given by
\begin{equation}
	\label{A.10}
	\begin{split}
		\mathcal{H} &=\frac{\mathcal{Q}_1^2}{2\tilde{C}_1}+\frac{\mathcal{Q}_2^2}{2\tilde{C}_2}
		+\frac{\Phi_1^2}{2\tilde{L}_1(\Phi_{x})}+\frac{\Phi_2^2}{2\tilde{L}_2(\Phi_{x})}
		+\frac{\mathcal{Q}_1\mathcal{Q}_2}{\tilde{C}_{12}}+\frac{\mathcal{Q}_1\mathcal{Q}_g}{\tilde{C}_{1g}}
		+\frac{\mathcal{Q}_2\mathcal{Q}_g}{\tilde{C}_{2g}}+\frac{\Phi_1\Phi_2}{\tilde{L}_{12}(\Phi_{x})},
	\end{split}
\end{equation}
where we have used $\tilde{\mathcal{Q}}_1=\mathcal{Q}_1+C_gV_g$, $\tilde{\mathcal{Q}}_{2}=\mathcal{Q}_2$ and defined the following dressed circuit parameters
\begin{flalign}
	\label{A13}
	\begin{split}
		&\tilde{C}_1=\tilde{C}_{1g}=\frac{(C_2+C_g)(C_1+C_g)+C_cC_2}{C_2+C_c},
		\tilde{C}_2=\frac{(C_2+C_g)(C_1+C_g)+C_cC_2}{C_c+C_1+C_g},  \tilde{L}_i(\Phi_{x})=\frac{L_i^2}{L_i+\frac{\Phi_0^2\alpha_i^2}{8\pi^2E_J\cos(\varphi_x)}},\\&
		\tilde{C}_{12}=\tilde{C}_{2g}=\frac{(C_2+C_g)(C_1+C_g)+C_cC_2}{C_c},
		\tilde{L}_{12}(\Phi_{x})=\frac{4\pi ^2E_J\cos(\varphi_x)L_1 L_2}{\alpha_1\alpha_2\Phi_0^2}.
	\end{split}
\end{flalign}

\subsubsection{Quantization of the Hamiltonian for a single microwave quantum memristor}
\label{A.2}
To proceed with the quantum mechanical description of the system, we promote the classical variables to quantum operators using $ \hat{Q}_\ell=2e\hat{n}_\ell$, $\hat{\Phi}_\ell=(\hat{\varphi}_\ell/2\pi)\Phi_0$, where $\hat{n}_{\ell}$ and $\hat{\varphi}_{\ell}$ are the cooper pair charge and phase operators, respectively, satisfying [$\hat{\Phi}_\ell,\hat{Q}_{\ell'}] = i\hbar\delta_{\ell,\ell'}$. Then, the Hamiltonian in Eq.~(\ref{A.10}) can be written as
\begin{equation}\label{18}
	\begin{split}
		\hat{\mathcal{H}} &=\sum_{\ell=1,2}\bigg[4E_{C\ell}\hat{n}_\ell^2 +\frac{E_{L\ell}(\Phi_{x})}{2}\hat{\varphi}_\ell^2+8E_{C\ell g}\hat{n}_{\ell}n_{g}\bigg]+8E_{C12}\hat{n}_1\hat{n}_2+E_{L12}(\Phi_{x})\hat{\varphi}_1\hat{\varphi}_2 ,
	\end{split}
\end{equation}
where $n_g =\mathcal{Q}_g/2e$ is the dimensionless gate charge, $E_{C\ell} = e^2/2\tilde{C}_\ell$ and $E_{L\ell}(\Phi_{x})=\Phi_0^2/(4\pi^2\tilde{L}_\ell(\Phi_{x}))$ are the charge and inductive energies, respectively, while $E_{C12}=e^2/2\tilde{C}_{12}$ and $E_{L12}(\Phi_{x}) =\Phi_0^2/(4\pi^2\tilde{L}_{12}(\Phi_{x}))$ are the coupling energies. Finally, we define the coupling energy of the $l$th resonator with the gate voltage as $E_{C\ell g}=e^2/2\tilde{C}_{\ell g}$. The charge and phase operators can be written in terms of the annihilation and creation operators, $\hat{n}_\ell=i n_\ell (\hat{a}_\ell^{\dagger}-\hat{a}_\ell)$ and $\hat{\varphi}_\ell = \varphi_\ell(\hat{a}_\ell^{\dagger}+\hat{a}_\ell) $, where $n_\ell = (E_{L\ell}/32E_{C\ell})^{1/4}$, $\varphi_\ell = (2E_{C\ell}/E_{L\ell})^{1/4}$ correspond to zero point fluctuations, leading to the following quantum Hamiltonian
\begin{eqnarray}\label{19}
	\begin{split}
		\hat{\mathcal{H}} &=\sum_{\ell=1,2}\bigg[\hbar\omega_\ell(\Phi_{x})\hat{a}^{\dagger}_\ell\hat{a}_\ell+iG_{ g\ell}(\Phi_{x},t)(\hat{a}_\ell^{\dagger}-\hat{a}_\ell)\bigg]
		+\lambda^{-}(\Phi_{x})(\hat{a}_1^{\dagger}\hat{a}_2^{\dagger}+\hat{a}_1\hat{a}_2)
		+\lambda^{+}(\Phi_{x})(\hat{a}_1^{\dagger}\hat{a}_2+\hat{a}_1\hat{a}_2^{\dagger}),
	\end{split}
\end{eqnarray}
where $\omega_\ell(\Phi_{x})=\sqrt{8E_{C\ell}E_{L\ell}(\Phi_{x})}/\hbar$ is the frequency of the $\ell$th resonator. Also, $G_{g\ell}(\Phi_{x},t)$ corresponds to the coupling strength with the external time-dependent gate voltage. Moreover, $\lambda^{\pm}(\Phi_{x})=I_{12}(\Phi_{x})\pm G_{12}(\Phi_{x})$ is the effective coupling strength where $I_{12}(\Phi_{x})$ and $G_{12}(\Phi_{x})$ are the tunable inductive and capacitive coupling strengths, respectively. They are expressed as
\begin{eqnarray*}
G_{ g\ell}(\Phi_{x},t) & = & 8E_{C\ell g}n_g\bigg(\frac{E_{L\ell}(\Phi_{x})}{32E_{C\ell}}\bigg)^{1/4} , \\ G_{12}(\Phi_{x}) & = & 2E_{C12}\bigg(\frac{E_{L1}(\Phi_{x})E_{L2}(\Phi_{x})}{4E_{C1}E_{C2}}\bigg)^{1/4} , \\ I_{12}(\Phi_{x}) & = & E_{L12}\bigg(\frac{4E_{C1}E_{C2}}{E_{L1}(\Phi_{x})E_{L2}(\Phi_{x})}\bigg)^{1/4}.
\end{eqnarray*}

The circuit and system parameters obtained after constrained optimization based on the mentioned approximations are summarized in the following Table \ref{tab:Table1},
\begin{table*}[ht]
	\centering
	\caption{Optimal circuit parameters.}
	\label{tab:Table1}
	\begin{tabular}{|llllllll|}
		\hline
		\multicolumn{8}{|c|}{Circuit Parameters} \\ \hline
		\multicolumn{1}{|l|}{$C_{c}$ {[}fF{]}} & \multicolumn{1}{l|}{$C_{1}$ {[}fF{]}} & \multicolumn{1}{l|}{$C_{2}${[}fF{]}} & \multicolumn{1}{l|}{$C_{J}$ {[}fF{]}} & \multicolumn{1}{l|}{$C_{g}$ {[}fF{]}} & \multicolumn{1}{l|}{$L_{1}$ {[}pH{]}} & \multicolumn{1}{l|}{$L_{2}$ {[}pH{]}} & $E_{J}/2\pi$ {[}GHz{]} \\ \hline
		\multicolumn{1}{|l|}{5.657} & \multicolumn{1}{l|}{413.5} & \multicolumn{1}{l|}{530.4} & \multicolumn{1}{l|}{536} & \multicolumn{1}{l|}{116.9} & \multicolumn{1}{l|}{746.2} & \multicolumn{1}{l|}{749.8} & 219.1 \\
 \hline
		\multicolumn{8}{|c|}{System parameters} \\ \hline
		\multicolumn{1}{|l|}{$\omega_{1}$ {[}GHz{]}} & \multicolumn{1}{l|}{$\omega_{2}$ {[}GHz{]}} & \multicolumn{1}{l|}{$\omega_{S}$ {[}GHz{]}} & \multicolumn{1}{l|}{$G_{12}/\omega_1$ } & \multicolumn{1}{l|}{$I_{12}/\omega_1$ } & \multicolumn{1}{l|}{$Z_{1}/Z_S$ } & \multicolumn{1}{l|}{$Z_{2}/Z_S$ } & $\omega_{S}/\omega_1$  \\ \hline
		\multicolumn{1}{|l|}{5} & \multicolumn{1}{l|}{5} & \multicolumn{1}{l|}{50} & \multicolumn{1}{l|}{0.005} & \multicolumn{1}{l|}{0.00005} & \multicolumn{1}{l|}{9.999} & \multicolumn{1}{l|}{9.999} & 10  \\
\hline
	\end{tabular}
\end{table*}

\subsection{Derivation of the circuit Hamiltonian for coupled microwave quantum memcapacitors}
\label{AppB}
In this section, we derive the quantum Hamiltonian of the coupled microwave quantum memcapacitors of Fig.~(\ref{Fig:coupled circuit_model}).

\subsubsection{Classical Hamiltonian for coupled microwave quantum memcapacitors}
The coupled system is described by the Lagrangian
\begin{flalign}\label{B1}
	\mathcal{L}=&
	\frac{ C_1}{2} \dot{\Phi} _1^2 +\frac{ C_3}{2} \dot{\Phi} _4^2-\frac{\Phi_1^2}{2L_1} -\frac{\Phi_4^2}{2L_3}
	+\frac{C_c }{2} \big\{(\dot{\Phi} _1-\dot{\Phi} _2)^2 + (\dot{\Phi} _2-\dot{\Phi} _3)^2 \big\}+\frac{C_2}{2}(\dot{\Phi} _2-\dot{\Phi} _3)^2 + \frac{C_4}{2}(\dot{\Phi} _5-\dot{\Phi} _6)^2-\frac{(\Phi_2-\Phi_3)^2}{2L_2} \nonumber\\&
	-\frac{(\Phi_5-\Phi_6)^2}{2L_4}
	+\frac{C_J}{2}(\dot{\Phi} _3^2 + \dot{\Phi} _6^2) +2E_J\cos(\varphi_x)\big\{\cos(\varphi_3) + \cos(\varphi_6) \big\}
	+ \frac{C_m}{2} (\dot{\Phi} _2-\dot{\Phi} _4)^2+\frac{C_g}{2} (\dot{\Phi} _1-V_g)^2,
\end{flalign}
where $\varphi_3= 2\pi\Phi_3/\Phi_0$ and $\varphi_6= 2\pi\Phi_6/\Phi_0$ are the superconducting phases in the respective devices, while $\varphi_x = 2\pi\Phi_x/\Phi_0$ is the external flux through the SQUID. Using the relation $ \mathcal{Q}_n=\partial \mathcal{L}/{\partial \dot{\Phi}_n}$, we obtain the relation between the node charges as
\begin{equation}
	\begin{split}
		&\mathcal{Q}_1 = (C_g+C_1+C_c)\dot{\Phi}_1-C_c\dot{\Phi}_2-C_gV_g,\quad
		\mathcal{Q}_2 =-C_c\dot{\Phi}_1+(C_c+C_m+C_2)\dot{\Phi}_2-C_2\dot{\Phi}_3-C_m\dot{\Phi}_4,\\&
		\mathcal{Q}_3 = -C_2\dot{\Phi}_2 + (C_J+C_2)\dot{\Phi}_3,\quad
		\mathcal{Q}_4 = -C_m\dot{\Phi}_2-(C_c+C_m+C_3)\dot{\Phi}_4-C_c\dot{\Phi}_5,\quad \\
		&\mathcal{Q}_5 = -C_c\dot{\Phi}_4-(C_c+C_4)\dot{\Phi}_5-C_4\dot{\Phi}_6,\quad
		\mathcal{Q}_6 = -C_4\dot{\Phi}_5+(C_J+C_4)\dot{\Phi}_6.
		\label{B2}
	\end{split}
\end{equation}
Similar to the previous section, we define $\tilde{\mathcal{Q}}_1=\mathcal{Q}_1+C_gV_g$, $\tilde{\mathcal{Q}}_2=\mathcal{Q}_2$,  $\tilde{\mathcal{Q}}_3=\mathcal{Q}_3$,
$\tilde{\mathcal{Q}}_4=\mathcal{Q}_4$,
$\tilde{\mathcal{Q}}_5=\mathcal{Q}_5$, and
$\tilde{\mathcal{Q}}_6=\mathcal{Q}_6$. The set of relations given in Eq.~(~\ref{B2}) can be written as $\vec{\tilde{\mathcal{Q}}} = \hat{C}\vec{\dot{\Phi}}$. By using the Legendre transformation $\mathcal{H}= \sum_n\mathcal{Q}_n\dot{\Phi}_n-\mathcal{L}$, we get the circuit Hamiltonian
\begin{flalign}
	\begin{split}
		\mathcal{H}_{2M} =& \sum_{i,j = 1(i\neq j)}^6 \frac{1}{2}\big( \tilde{Q}_i^2 C^{-1}_{i,i}
				+\tilde{Q}_i \tilde{Q}_j C^{-1}_{i,j} \big)
				+\frac{\Phi_1^2}{2L_1}+\frac{(\Phi_2-\Phi_3)^2}{2L_2}+\frac{\Phi_4^2}{2L_3}+\frac{(\Phi_5-\Phi_6)^2}{2L_4}\\&-2E_J\cos(\varphi_x)\big\{\cos(\varphi_3)+\cos(\varphi_6)\big\},
		\label{B4}
	\end{split}
\end{flalign}
where $C_{ij}^{-1}$ corresponds to the inverse capacitance matrix element of $\hat{C}$. As with a single microwave quantum memcapacitor, we consider the low-impedance regime of the SQUIDs $(\dot{\Phi}_3\ll \dot{\Phi}_{1(2)}, \dot{\Phi}_6 \ll \dot{\Phi}_{4(5)})$ obtaining the relation between the node charges
\begin{equation}
	\begin{split}
		\tilde{\mathcal{Q}}_3= -C_2(B_{21}^{-1}\tilde{\mathcal{Q}}_1+B_{22}^{-1}\tilde{\mathcal{Q}}_2+B_{23}^{-1}\tilde{\mathcal{Q}}_4+B_{24}^{-1}\tilde{\mathcal{Q}}_5),\quad
		\tilde{\mathcal{Q}}_6= -C_4(B_{41}^{-1}\tilde{\mathcal{Q}}_1+B_{42}^{-1}\tilde{\mathcal{Q}}_2+B_{43}^{-1}\tilde{\mathcal{Q}}_4+B_{44}^{-1}\tilde{\mathcal{Q}}_5) ,
		\label{B8}
	\end{split}
\end{equation}
where $B^{-1}_{ij}$ are the elements of the inverse of matrix $\hat{B}$ given by
\begin{eqnarray}
	\hat{B} = \begin{pmatrix}
		C_c+C_g+C_1 & -C_c & 0 & 0  \\
		-C_c & C_c+C_m+C_2 & -C_m & 0  \\
		0 & -C_m & C_c+C_m+C_3 & -C_c  \\
		0 & 0 & -C_c & C_c+C_4 \\
	\end{pmatrix}.
	\label{Eq.B5}
\end{eqnarray}
Similarly, for the node fluxes, using the Euler-Lagrange equation $\partial \mathcal{L}/\partial\Phi_i - d(\partial\mathcal{L}/\partial \dot{\Phi}_i)/dt = 0$ and considering second linearized regime of the Junctions ($\sin(\varphi_{3(6)})=\varphi_{3(6)}$) and low-impedance regime of the SQUIDs ( $\ddot{\Phi}_3\ll\ddot{\Phi}_{1(2)}$, $\ddot{\Phi}_6\ll\ddot{\Phi}_{4(5)}$), we get
\begin{eqnarray}
	\varphi_{3(6)}= \frac{(-C_{2(4)})\Phi_0}{4\pi E_J\cos(\varphi_x)}\bigg(B_{2(4)1}^{-1}\frac{\Phi_1}{L_1}+B_{2(4)2}^{-1}\frac{\Phi_2}{L_2}+B_{2(4)3}^{-1}\frac{\Phi_4}{L_3}+B_{2(4)4}^{-1}\frac{\Phi_5}{L_4}\bigg) . \nonumber \\
	\label{B10}
\end{eqnarray}
Furthermore, we express the Hamiltonian in terms of Eq.~(\ref{B10}) considering high plasma frequency( $\Phi_3\ll\Phi_{1(2)}$, $\Phi_6\ll\Phi_{4(5)}$) and approximating $\cos(\varphi_{3(6)}) \approx 1 -\varphi_{3(6)}^2 $. Using Eq.~(\ref{B8}) and Eq.~(\ref{B10}) in Eq.~(\ref{B4}), we obtain
\begin{equation}\label{B11}
	\mathcal{H}_{2M} = \sum_{i=1}^{2}\mathcal{H}_{i}+\mathcal{H}_{c} ,
\end{equation}
where $\mathcal{H}_{i}$ is the Hamiltonian of the single microwave quantum memcapacitor derived in the last section, Eq. (\ref{A.10}), and the coupling Hamiltonian is
\begin{eqnarray*}
\mathcal{H}_{c} = \frac{\mathcal{Q}_1\mathcal{Q}_4}{\tilde{C}_{13}}+\frac{\mathcal{Q}_1\mathcal{Q}_5}{\tilde{C}_{14}}+\frac{\mathcal{Q}_2\mathcal{Q}_4}{\tilde{C}_{23}}+\frac{\mathcal{Q}_2\mathcal{Q}_5}{\tilde{C}_{24}}+\frac{\Phi_1\Phi_4}{\tilde{L}_{13}}+\frac{\Phi_1\Phi_5}{\tilde{L}_{14}}+\frac{\Phi_2\Phi_4}{\tilde{L}_{23}}	+\frac{\Phi_2\Phi_5}{\tilde{L}_{24}} .
\end{eqnarray*}
Here, we have used $\tilde{\mathcal{Q}}_1=\mathcal{Q}_1+C_gV_g$, $\tilde{\mathcal{Q}}_2=\mathcal{Q}_2$, $\tilde{\mathcal{Q}}_3=\mathcal{Q}_3$, the effective coupling capacitances are defined as
\begin{flalign*}
\nonumber
\tilde{C}_{m,n(m=1,2;n=3,4)} = &\bigg(C^{-1}_{m,n+1}+C^{-1}_{3,3}C_2^2B_{2,m}^{-1}b_{2,n}^{-1}-C_2C^{-1}_{m,3}B_{2n}^{-1}-C_2C^{-1}_{3,n+1}B_{2,m}^{-1}+C_4^2C^{-1}_{6,6}B_{4,m}^{-1}B_{4,n}^{-1}
-C_4C^{-1}_{m,6}B_{4,n}^{-1}\\&-C_4C_{n+1,6}B_{4,m}^{-1}+C_4 C_2 C_{3,6}^{-1}B_{2,m}^{-1}B_{4,n}^{-1}+C_4 C_2 C_{3,6}^{-1}B_{2,n}^{-1}B_{4,m}^{-1}\bigg)^{-1} ,
\end{flalign*}
and the effective coupling inductances reads
\begin{flalign*}
\nonumber
&
\tilde{L}_{m,n(m=1,2,n=3,4)}=\frac{ 4\pi E_J\cos(\varphi_x)}{\Phi_0^2}\bigg(\frac{{C_2}^2B_{2,m}^{-1}B_{2,n}^{-1}}{L_mL_n}+\frac{{C_4}^2B_{4,m}^{-1}B_{4,n}^{-1}}{L_mL_n}\bigg)^{-1}.
\end{flalign*}

\subsubsection{Quantization of the Hamiltonian for coupled microwave quantum memcapacitors}
\label{App.B2}
By promoting the charge and phase variables to quantum operators, using $ \hat{Q}_\ell=2e\hat{n}_\ell$, $\hat{\Phi}_\ell=(\hat{\varphi}_\ell/2\pi)\Phi_0$ to satisfy the canonical commutation relation [ $\hat{\Phi}_\ell,\hat{Q}_{\ell'}] = i\hbar\delta_{\ell,\ell'}$, we obtain the quantum Hamiltonian of the coupled microwave quantum memcapacitors as
\begin{equation}
\hat{\mathcal{H}}_{2M} = \sum_{i=1}^{2}\hat{\mathcal{H}}_{i}+\hat{\mathcal{H}}_c .
\end{equation}
Here, $\hat{\mathcal{H}}_{i}$ is the quantum Hamiltonian of a single microwave quantum memcapacitor (see Eq. (\ref{18})) and the quantized coupling Hamiltonian $\hat{\mathcal{H}}_c$ reads
\begin{eqnarray}
&\hat{\mathcal{H}}_c=\sum_{n=1(2),m=3(4)}\big(4E_{Cnm}\hat{n}_n\hat{n}_{m+1} +E_{Lnm}\hat{\varphi}_n\hat{\varphi}_{m+1}\big),
\end{eqnarray}
where the coupling capacitance and inductance energies are $ E_{Cnm(n=1,2,m=3,4)}=e^2/(2\tilde{C}_{nm})$, $E_{Lnm(n=1,2,m=3,4)}(\Phi_{x}) =\Phi_0^2/(4\pi^2\tilde{L}_{nm}(\Phi_{x}))$.
We express the charge and phase operators in terms of the annihilation and creation operators
\begin{eqnarray*}
	&&\hat{n}_{1(2)}=i n_{1(2)} (\hat{a}_{1(2)}^{\dagger}-\hat{a}_{1(2)}),\quad  \hat{\varphi}_{1(2)} = \varphi_{1(2)}(\hat{a}_{1(2)}^{\dagger}+\hat{a}_{1(2)}),\\&&
 \hat{n}_{4(5)}=i n_{4(5)} (\hat{b}_{1(2)}^{\dagger}-\hat{b}_{1(2)}),\quad
	\hat{\varphi}_{4(5)} = \varphi_{4(5)}(\hat{b}_{1(2)}^{\dagger}+\hat{b}_{1(2)}) ,
\end{eqnarray*}
where  $n_{1(2)} = (E_{L1(2)}/32E_{C1(2)})^{1/4}$, $\varphi_{1(2)} = (2E_{C1(2)}/E_{L1(2)})^{1/4}$,  $n_{4(5)} = (E_{L3(4)}/32E_{C3(4)})^{1/4}$, $\varphi_{4(5)} = (2E_{C3(4)}/E_{L3(4)})^{1/4}$ with $E_{C\ell=1,2,3,4} = e^2/2\tilde{C}_{\ell}$,
$E_{L\ell=1,2,3,4}(\Phi_{x})=\Phi_0^2/(4\pi^2\tilde{L}_\ell(\Phi_{x}))$. Then, we obtain the second quantized Hamiltonian
\begin{flalign*}
&\hat{\mathcal{H}}_{1} =	
\sum_{\ell=1,2}\bigg[\omega_{\ell}(\Phi_{x})\hat{a}_\ell^{\dagger}\hat{a}_\ell+i{G}_{ g\ell}(\Phi_x,t) (\hat{a}_\ell^{\dagger}-\hat{a}_\ell)\bigg]+\lambda^{+}(\Phi_{x})(\hat{a}_1^{\dagger}\hat{a}_2+\hat{a}_1\hat{a}_2^{\dagger})
+\lambda^{-}(\Phi_{x})(\hat{a}_1^{\dagger}\hat{a}_2^{\dagger}+\hat{a}_1\hat{a}_2),\\&
\hat{\mathcal{H}}_{2}=	
\sum_{\ell=1,2}\bigg[\Omega_{\ell}(\Phi_{x})\hat{b}_\ell^{\dagger}\hat{b}_\ell
+i{J}_{ g\ell}(\Phi_x,t) (\hat{b}_\ell^{\dagger}-\hat{b}_\ell)  \bigg]
+\Lambda^{+}(\Phi_{x})(\hat{b}_1^{\dagger}\hat{b}_2+\hat{b}_1\hat{b}_2^{\dagger})
+\Lambda^{-}(\Phi_{x})(\hat{b}_1^{\dagger}\hat{b}_2^{\dagger}+\hat{b}_1\hat{b}_2),\\&
\hat{\mathcal{H}}_{c}			=\sum_{j,k=1}^{2}\bigg[\gamma^{+}_{j,k}(\Phi_{x})(\hat{a}_j^{\dagger}\hat{b}_k+\hat{a}_j\hat{b}_k^{\dagger})
+\gamma^{-}_{j,k}(\Phi_{x})(\hat{a}_j^{\dagger}\hat{b}_k^{\dagger}+\hat{a}_j\hat{b}_k)\bigg],
\end{flalign*}
where $\omega_{1(2)}(\Phi_{x})=\sqrt{8E_{C 1(2)}E_{L1(2)}(\Phi_{x})}/\hbar$ and $\Omega_{1(2)}(\Phi_{x})=\sqrt{8E_{C3(4)}E_{L3(4)}(\Phi_{x})}/\hbar$ are the frequencies of the resonator in each microwave quantum memcapacitor. Also, $G_{{ g\ell}}(\Phi_x,t)$ and $J_{ g\ell}(\Phi_x,t)$ corresponds to the coupling strength between the resonators with gate voltage. Here,
$\lambda^{\pm}(\Phi_{x})={I}_{12}(\Phi_{x})\pm {G}_{12}(\Phi_{x}) $, and $\Lambda^{\pm}(\Phi_{x})={F}_{12}(\Phi_{x})\pm {J}_{12}(\Phi_{x}) $ are the effective coupling strengths of each single microwave quantum memcapacitor, while $\gamma^{\pm}(\Phi_{x})={K}_{j,k}(\Phi_{x})\pm{M}_{j,k}(\Phi_{x})$ are the effective coupling strength between oscillators of the two microwave quantum memcapacitor defined as
\begin{align*}
{K}_{j=1,2,k=3,4}(\Phi_{x})= E_{Ljk}(\Phi_{x})\bigg(\frac{4E_{Cj}E_{Ck}}{E_{Lj}(\Phi_{x})E_{Lk}(\Phi_{x})}\bigg)^{1/4},&&
{M}_{j=1,2,k=3,4}(\Phi_{x})= 2E_{Cjk}\bigg(\frac{E_{Lj}(\Phi_{x})E_{Lk}(\Phi_{x})}{4E_{Cj}E_{Ck}}\bigg)^{1/4}.
\end{align*}
Finally, we summarize the coupled systeme parameters used in the main text, where Table \ref{tab:Table2} shows the optimal case.
\begin{table*}[ht]
	\centering
	\setlength\tabcolsep{5pt}
	\caption{Coupled device circuit parameters.}
	\label{tab:Table2}
	\begin{tabular}{|llllllll|}
		\hline
		\multicolumn{8}{|c|}{Circuit Parameters} \\ \hline
		\multicolumn{1}{|l|}{$C_{c}$ {[}fF{]}} & \multicolumn{1}{l|}{$C_{1}$ {[}fF{]}} & \multicolumn{1}{l|}{$C_{2}${[}fF{]}}& \multicolumn{1}{l|}{$C_{3}$ {[}fF{]}} & \multicolumn{1}{l|}{$C_{4}${[}fF{]}}  & \multicolumn{1}{l|}{$C_{J}$ {[}fF{]}} & \multicolumn{1}{l|}{$C_{g}$ {[}fF{]}}&	\multicolumn{1}{l|}{$C_{m}${[}fF{]}}
		\\
		\hline
		\multicolumn{1}{|l|}{5.657} & \multicolumn{1}{l|}{413.5} & \multicolumn{1}{l|}{530.4} &   \multicolumn{1}{l|}{413.5} & \multicolumn{1}{l|}{530.4}&
		\multicolumn{1}{l|}{536.0} & \multicolumn{1}{l|}{116.9} 	&
		\multicolumn{1}{l|}{11.69}\\ \hline	\multicolumn{1}{|l|}{$L_{1}$ {[}pH{]}} & \multicolumn{1}{l|}{$L_{2}$ {[}pH{]}}
		&\multicolumn{1}{l|}{$L_{3}$ {[}pH{]}} & \multicolumn{1}{l|}{$L_{4}$  {[}pH{]}} &\multicolumn{1}{l|} {$E_{J}/2\pi$ {[}GHz{]}}  &  & & \\ \hline
		\multicolumn{1}{|l|}{746.2} & \multicolumn{1}{l|}{749.8}&\multicolumn{1}{l|}{746.2}& \multicolumn{1}{l|}{749.8} & \multicolumn{1}{l|}{219.1}  &  & &
		\\ \hline	
		\multicolumn{8}{|c|}{System parameters} \\ \hline
		\multicolumn{1}{|l|}{$\omega_{1}$ {[}GHz{]}} & \multicolumn{1}{l|}{$\omega_{2}$ {[}GHz{]}} &\multicolumn{1}{l|}{$\Omega_{1}$ {[}GHz{]}} & \multicolumn{1}{l|}{$\Omega_{2}$ {[}GHz{]}}&  	\multicolumn{1}{l|}{$\lambda^{+}/\omega_1$ } & \multicolumn{1}{l|}{$\lambda^{-}/\omega_1$ }
		&\multicolumn{1}{l|}{$\Lambda^{+}/\omega_1$ }&
		\multicolumn{1}{l|}{$\Lambda^{-}/\omega_1$ }
		\\ \hline
		\multicolumn{1}{|l|}{5} & \multicolumn{1}{l|}{4.97} & \multicolumn{1}{l|}{5.579} & \multicolumn{1}{l|}{5.034} &
		\multicolumn{1}{l|}{0.00569} &
		\multicolumn{1}{l|}{-0.00529} & \multicolumn{1}{l|}{0.00652} &
		\multicolumn{1}{l|}{-0.00597}	
		\\ \hline
		\multicolumn{1}{|l|}{$\gamma^{+}_{1,1}\omega_1$ }	&\multicolumn{1}{l|}{$\gamma^{-}_{1,1}/\omega_1$ }&
		\multicolumn{1}{l|}{$\gamma^{+}_{1,2}/\omega_1$ } &\multicolumn{1}{l|}{$\gamma^{-}_{1,2}/\omega_1$} & \multicolumn{1}{l|}{$\gamma^{+}_{2,1}/\omega_1$ } &\multicolumn{1}{l|}{$\gamma^{-}_{2,1}/\omega_1$} & \multicolumn{1}{l|}{$\gamma^{+}_{2,2}/\omega_1$}& \multicolumn{1}{l|}{$\gamma^{-}_{2,2}/\omega_1$}  \\ \hline
		\multicolumn{1}{|l|}{0.000145}&
		\multicolumn{1}{l|}{-0.000134}&\multicolumn{1}{l|}{0.0} &\multicolumn{1}{l|}{0.0}
		& \multicolumn{1}{l|} {0.0139}& \multicolumn{1}{l|}{-0.0128}
		&\multicolumn{1}{l|}{0.00016} & \multicolumn{1}{l|}{-0.00014}
		\\ \hline	
	\end{tabular}
\end{table*}

\section{Data Availability}
All the data that support this work is available under a proper request to the corresponding author.

\section{Author Contributions}
X.-Y. Qiu and S. Kumar were in charge of the circuit quantization, and numerical calculation of quantum entanglement. F. A. C\'ardenas-L\'opez was in charge of the final production of the different hysteresis loops and final form of the results. G. A. Barrios was in charge of the memory features interpretations as well as of the proper discussion of the results from a memristive point of view. E. Solano and F. Albarr\'an-Arriagada supervised the work and propose the superconducting circuit design as possible memory device. All the authors contribute to the writting and final version of this work, as well as, all the authors review and approve this version.
\\

\underline{\textbf{Competing Interests}.}- The authors declare no competing interests

\end{document}